\documentclass[10pt, a4paper]{article}   	
\usepackage{geometry}  
\usepackage{mathrsfs}   
\usepackage{amsmath,amsthm, amssymb, latexsym}
\usepackage{mathtools} 
\usepackage{bbold}          		
\usepackage{graphicx}				

\usepackage[linesnumbered,ruled]{algorithm2e}
\usepackage{color}
\usepackage{csquotes}
\usepackage{tabularx}
\usepackage{float}

\usepackage{hyperref}
 \hypersetup{
    colorlinks=true,
    linkcolor=blue,
    filecolor=magenta,      
    urlcolor=cyan,
}

\newtheoremstyle{mydefi}
 {15pt}
  {15pt}
  {}
  {}
  {\bfseries}
  {:}
  {.5em}
  {}
\newtheoremstyle{mytheo}
  {15pt}
  {15pt}
  {\slshape}
  {}
  {\bfseries}
  {:}
  {.5em}
  {}

\theoremstyle{mytheo}
\newtheorem{theorem}{Theorem}[section]

\theoremstyle{mydefi}

\theoremstyle{mydefi}

\title{A comparative study of estimation methods in quantum tomography }

\author{Anirudh Acharya, Theodore Kypraios  and M\u{a}d\u{a}lin Gu\c{t}\u{a} \\[4mm]
School of Mathematical Sciences, University of Nottingham,\\ University Park, NG7 2RD Nottingham, UK}

\date{}							

\begin{document}
\maketitle

\begin{abstract}
As quantum tomography is becoming a key component of the quantum engineering toolbox, there is a need for a deeper understanding of the multitude of estimation methods available. 
Here we investigate and compare several such methods: maximum likelihood, least squares, generalised least squares, positive least squares, thresholded least squares and projected least squares. The common thread of the analysis is that each estimator projects the measurement data onto a parameter space with respect to a specific metric, thus allowing us to study the relationships between different estimators. 

The asymptotic behaviour of the least squares and the projected least squares estimators is studied in detail for the case of the covariant measurement and a family of states of varying ranks. This gives insight into the rank-dependent risk reduction for the projected estimator, and uncovers an interesting non-monotonic behaviour of the Bures risk. These asymptotic results complement recent non-asymptotic concentration bounds of \cite{GutaKahnKungTropp} which point to strong optimality properties, and high computational efficiency of the projected linear estimators.

To illustrate the theoretical methods we present results of an extensive simulation study. An app running the different estimators has been made available online.

\end{abstract}

\section{Introduction}

The problem of estimating unknown parameters of quantum systems has been at the forefront of early investigations into the statistical nature of quantum information \cite{Belavkin1976,Holevo,Helstrom69,Yuen&Lax,BraunsteinCaves}. Traditionally, key research topics have been the design of optimal measurements and estimation procedures \cite{Massar&Popescu,Vidal,D'Ariano.0,Gill&Massar,Keyl&Werner,Bagan&Baig&Tapia,Bagan&Gill,Hayashi&Matsumoto,Bagan&Gill}, and theoretical aspects of quantum Fisher information and asymptotic estimation \cite{Barndorff-Nielsen&Gill&Jupp,Fujiwara&Nagaoka,Matsumoto,Hayashi&Matsumoto,Gill&Guta&Artiles,Petz&Jencova,KahnGuta2009,GillGuta2013}, see also the monographs \cite{ParisRehacek,Leonhardt,Hayashi.editor}. 

More recently, quantum tomography has become a crucial validation tool current quantum technology applications \cite{Haffner2005,Monz2011,tenqubit2010,SchwemmerToth}. The experimental challenges have stimulated research in new directions such as compressed sensing \cite{CSnoRIP,Liu2011,CSerrorbounds,SteffensRio,Cramer:2010,AcharyaTheoGuta,AcharyaGuta}, estimation of permutationally invariant states \cite{TobiasToth}, adaptive and selflearning tomography \cite{MahlerRozema,GranadeFerrieFlammia2017,PereiraDelgado2018,
Ferrie2014,HannemannReiss,QiBo}, incomplete tomography \cite{Hradilincomplete}, 
minimax bounds \cite{FerrieBK,AcharyaBures}, Bayesian estimation \cite{BlumeKohout,GranadeCombes}, and confidence regions \cite{Audenaert_Scheel,christandl2012,SuessGross,FaistRenner,LiEnglert2016}. Since `full tomography' becomes impossible for systems composed of even a moderate number of qubits, research has focused on the estimation of states which belong to smaller dimensional models which nevertheless capture relevant physical properties, such as low rank states \cite{KoltchinskiiXia,spectralthresholding,HaahHarrow2017,kueng_low_2017,GutaKahnKungTropp}
and matrix product states \cite{Cramer:2010,BaumgratzCramerPlenio,LanyonMaier}.


In this paper we analyse and compare several estimation methods for fixed (non-adaptive) measurement  designs, with a focus on risks (mean errors) for different loss functions, asymptotic theory, relationships between estimators and low rank behaviour. The measurement scenarios include repeated measurements with respect to Pauli bases (as customary in multiple qubits 
experiments), random bases measurements, and the covariant measurement.
The loss functions are given by the Frobenius, trace-norm, operator-norm, Bures and Hellinger distances. Each section deals with one class of estimators and the results of a comparative simulations study are presented at the end of the paper. While most estimators have been previously considered in the literature, our aim is to investigate them from a common perspective, as projections of data onto the parameter space. 
Another aim is to understand and quantify the reduction in statistical error between an unconstrained estimator such as least squares, and an estimator which take into account the physical properties of the parameter space, such as the projected least squares estimator. Among the original results, we derive the asymptotic error rates of these estimators on a class of low rank states in the covariant measurement scenario. Finally, we discuss the computational efficiency of different methods. 

\begin{figure}
 \centering
 \includegraphics[width = 14cm]{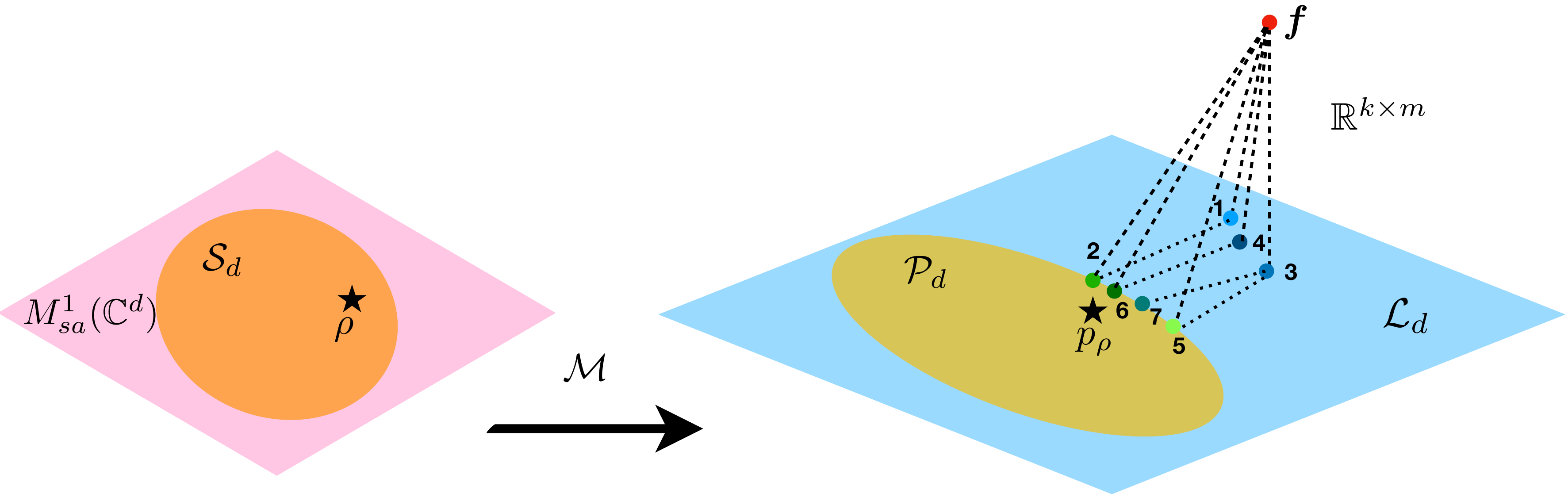}
 \caption{
A measurement $\mathcal{M}$ maps the set of states $\mathcal{S}_d$  onto outcome probabilities $\mathcal{P}_d$. Estimators are represented via their images through $\mathcal{M}$,
which are obtained by projecting the empirical frequencies $\boldsymbol{f}$ onto the hyperplane $\mathcal{L}_d=\mathcal{M}(M^1_{sa}(\mathbb{C}^d))$ or  the convex set $\mathcal{P}_d$ with respect to a metric. The  uML (1) and ML (2) estimators are projections with respect to relative entropy.  The GLS (4) and  posGLS (6) estimators are projections with respect to the covariance metric, and are asymptotically equivalent to the uML and ML.  The  LS (3) and posLS (5) estimators project $\boldsymbol{f}$ with respect to the euclidian distance. The PLS (7) estimator projects LS onto $\mathcal{P}_d$ with respect to the Frobenius distance inherited from $M(\mathbb{C}^d)$.
}\label{fig.big_picture}
 \end{figure}

Below we summarise the paper using Figure \ref{fig.big_picture} for illustration. A measurement $\mathcal{M}$ on $d$-dimensional system is a positive affine map from the convex set of states $\mathcal{S}_d$ (constrained parameter space) onto the space of outcome probabilities $\mathcal{P}_d\subset \mathbb{R}^z$, where $z$ is the number outcomes. The image of the set of trace-one selfadjoint matrices  $M^1_{sa}(\mathbb{C}^d)$ (unconstrained parameter space) is  a hyperplane $\mathcal{L}_d $ of $\mathbb{R}^z$, which contains 
$\mathcal{P}_d$. For informationally complete measurements the map $\mathcal{M}$ is injective and we can identify all matrix estimators (whether positive or not) with their images in $\mathcal{L}_d$.

In section \ref{sec1:Fisher} we review the use of Fisher information in asymptotic normality theory and discuss to what extent it is applicable to the study of the maximum likelihood (ML) estimator in quantum tomography \cite{MLparis,MLrobin,BlumeKohout_2018,Englert2011}. We distinguish between two ML estimators: the \emph{unconstrained} estimator 
$\hat{\rho}_{\rm uML}$, and the `physical' estimator $\hat{\rho}_{\rm ML}$. 
The former is the maximiser of the likelihood, seen as a function over $M^1_{sa}(\mathbb{C}^d)$, and may not be a density matrix. The latter performs the same optimisation but  restricted to the physical space of density matrices 
$\mathcal{S}_d \subset M^1_{sa}(\mathbb{C}^d)$. For large sample size, the unconstrained estimator is asymptotically normal, while the constrained estimator is equal to its projection onto $\mathcal{S}_d $ with respect to the metric defined by the classical Fisher information 
\cite{BlumeKohout_2018}.

In section \ref{sec.LS} we analyse the least squares (LS) estimator $\hat{\rho}_{\rm LS}$. This exploits the linear dependence between probabilities and states which translates into a linear regression problem 
$$
\boldsymbol{f} = X \boldsymbol{\beta} + \boldsymbol{\epsilon}
$$
where $\boldsymbol{f} $ denotes the vector of observed frequencies of measurement outcomes, 
$X$ is a fixed measurement design matrix, $\boldsymbol{\beta}$ is a vectorisation of the unknown state $\rho$, and $\boldsymbol{\epsilon}$ is the statistical noise. 
The LS estimator minimises the prediction error 
$\| X \hat{\boldsymbol{\beta}} -\boldsymbol{f} \|^2$ over all vectors $\hat{\boldsymbol{\beta}}$, 
and is the simplest and fastest estimator to compute. 
However, LS is significantly less accurate than ML, especially for low rank states and does not produce a physical state. Nevertheless, the LS has been the focus of recent investigations 
\cite{spectralthresholding,SugiyamaTurnerMurao,GutaKahnKungTropp}  as a first step towards more accurate estimators. Here we review some of the \emph{non-asymptotic} concentration bounds for the operator-norm error of the LS estimator. We then study the \emph{asymptotic} properties of the LS estimator in the context of covariant measurements, and rank $r$ states 
$\rho_r$ with equal non-zero eigenvalues. By exploiting the symmetry of the measurement we  compute the explicit expression of the risk with respect to the Frobenius distance.  Furthermore, we show that for large $d$, the eigenvalues of the `error matrix' $\hat{\rho}_{\rm LS} - \rho_r$ are  approximately distributed according to the Wigner semicircle law on the interval 
$[- 2\sqrt{d/N}, 2\sqrt{d/N}]$, cf. Figure \ref{fig.wigner}. This provides the asymptotic estimates of the operator-norm, and the trace-norm errors of $2\sqrt{d/N}$ and respectively $8 d^{3/2}/ (3\pi\sqrt{N})$, which complement the non-asymptotic bounds of \cite{GutaKahnKungTropp}.

In section \ref{sec.gls} we discuss the generalised least squares (GLS) estimator. Unlike the LS estimator which minimises the prediction error, 
the GLS aims to optimise the estimation error, e.g. $\mathbb{E}\|\hat{\boldsymbol{\beta}}- \boldsymbol{\beta} \|^2$ and needs to take into account the covariance matrix of the multinomial 
noise $\boldsymbol{\epsilon}$. We show that for large samplesize $N$, the GLS estimator is asymptotically normal and is equivalent to the uML estimator.

Section \ref{sec1:TLS} reviews the thresholded least squares (TLS) estimator proposed in \cite{spectralthresholding}. This is obtained by projecting the LS estimator onto the set of states whose non-zero eigenvalues are above a certain `noise level' chosen in a data-driven fashion. The projection can be computed efficiently in terms of the eigenvalues of the LS estimator, and the truncation leads to significant error reduction for low rank states. In practice, an additional improvement is achieved by using the GLS estimator as starting point.

Section \ref{sec.pos.ls} discusses the positive least squares (posLS) estimator which optimises the prediction error over the physical space of density matrices, rather than over the selfadjoint matrices as is the case for the LS estimator \cite{KalevBaldwin}.  This leads to higher accuracy, but its computational complexity is similar to that of ML. However, in the case of the covariant measurements we find that posLS is equivalent to the projected least squares estimator discussed below, which \emph{can} be computed efficiently. By restricting the GLS optimisation over density matrices we obtain the positive generalised least squares estimator (posGLS), which is shown to be asymptotically equivalent to the ML estimator.

Section \ref{sec.pls} deals with the projected least squares (PLS) estimator. This is obtained by projecting the LS estimator onto the space of density matrices with respect to the Frobenius distance \cite{SugiyamaTurnerMurao,GutaKahnKungTropp}. It is faster than TLS, and has similar statistical properties. In \cite{GutaKahnKungTropp} is was shown that the  PLS estimator satisfies rank-dependent concentration bounds and its trace-norm error scales 
as $O( r^2 \cdot d \cdot \log d  /\sqrt{N})$ for a broad class of 2-design measurements including mutually unbiased bases, stabiliser states, and symmetric informationally complete measurements,  and as $O(r^2 \cdot d^{1.6}\cdot \log d  /\sqrt{N})$ for Pauli bases measurements. In this paper we focus on the asymptotic behaviour of the PLS estimator in the case of covariant measurements. Inspired by the techniques developed in \cite{BlumeKohout_2018} we show how the random matrix properties of the LS can be used to derive the asymptotic Frobenius  and Bures risks of the PLS estimator for low rank states and 
large dimension $d$ and samplesize $N$. In particular we uncover an interesting behaviour of the Bures risk which (unlike the Frobenius or the trace norm risks) increases steeply with rank for low rank states and then decreases towards full rank, cf. Figure \ref{fig.asymptotics.LS.PLS}.

Section \ref{sec.simulations} presents the results of a comparative simulations study of the 
proposed estimators. We simulated data for a range of states of 3 and 4 atoms of different ranks, with different samplesizes, and measurement setups. For each choice we produced 100 datasets in order to estimate risks of different estimators. The measurements are chosen to be either Pauli bases measurements (as standard in ion trap experiments) or random basis measurements. The different estimators and their risks corresponding to Frobenius, Bures, Hellinger, and trace-norm distances were computed and the results are illustrated and discussed, cf. Figures \ref{fig1:frobeniuspauli}, \ref{fig1:frobeniusrandom4}, \ref{fig1:burespauli}, \ref{fig1:buresrandom3}, \ref{fig:mse.with.n}, \ref{fig:mf_3_random}, \ref{fig.L11}, and \ref{fig.L2}. A complete set of simulation results available online via an interactive Rshiny application at \url{https://rudhacharya.shinyapps.io/plots/}. We also created an online estimation app which computes the key estimators for a range of states of certain characteristics, or user 
uploaded states. This is available at \url{https://shiny.maths.nottingham.ac.uk/shiny/qt_dashboard/}.

\section{Quantum tomography}\label{sec.qt}

In quantum tomography, the aim is to estimate an unknown state from outcomes of measurements performed on an ensemble of identical prepared systems. Although a large part of our theoretical considerations hold in a general setting, we choose to formulate the tomography problem in the specific context of a system consisting of multiple qubits and projective measurements with respect to several bases. This keeps the discussion closer to realistic experimental procedures, and facilitates the understanding of our simulation results.

We consider composite systems consisting of $n$ qubits, with associated  Hilbert space $\mathbb{C}^d$, where $d=2^n$. The state $\rho$ belongs to the space of 
$d\times d$ density matrices  
$\mathcal{S}_d\subset M(\mathbb{C}^d)$. 
In our analysis we will distinguish between the constrained parameter space 
$\mathcal{S}_d$ whose elements are trace-one positive matrices, and the unconstrained parameter space $M^1_{sa}(\mathbb{C}^d)$ consisting of trace-one selfadjoint matrices. 
This will allow us to consider procedure which produce constrained, or unconstrained estimators.

The measurement strategies consist of performing standard, von Neumann projective measurements with respect to a number of orthonormal bases (ONB), which are chosen deterministically or randomly. In particular, we focus on two scenarios: Pauli basis measurements and measurements that are drawn randomly from the uniform measure over all ONBs. While the former setup is commonly employed in experiments \cite{Haffner2005}, the latter is less restrictive, more amenable to theoretical analysis, and can serve as a benchmark for current experiments. We also consider covariant measurements in the context of the asymptotic theory of the least squares and respectively projected least squares estimators in sections \ref{sec.asymptotics.ls.covariant} and \ref{sec.PLS.covariant}, and refer to these for further details.

 In the Pauli bases setup, one measures an arbitrary Pauli observable $\sigma_x, \sigma_y$, or 
 $\sigma_z$ on each of the $n$ ions simultaneously. Therefore, each measurement is labelled by a sequence $\boldsymbol{s} = (s_1, \ldots, s_n) \in \{ x, y, z\}^n$, and there are $3^n$ possible measurement bases. 
%
Such a measurement produces a $\pm1$ outcome from each ion, and we let $\boldsymbol{o} \in \{+1, -1 \}^n$ be the full record of outcomes from all the $n$ ions. The probability of obtaining a particular outcome $\boldsymbol{o}$ is given by $p_{\rho} (\boldsymbol{o}|\boldsymbol{s} ):= {\rm Tr}(\rho P_{\boldsymbol{o}}^{\boldsymbol{s}})$, where $P_{\boldsymbol{o}}^{\boldsymbol{s}}$ is the one-dimensional projection onto the tensor product of eigenvectors of Pauli matrices 
\begin{equation*}
P_{\boldsymbol{o}}^{\boldsymbol{s}} = \vert e^{s_1}_{o_1} \rangle \langle e^{s_1}_{o_1} \vert \otimes \ldots \otimes \vert e^{s_n}_{o_n} \rangle \langle e^{s_n}_{o_n} \vert , \quad \sigma_s |e^{s}_{o}\rangle = o |e^{s}_{o}\rangle.
\end{equation*}
More generally, the measurement design is defined by a collection $\mathscr{S} = \{ \boldsymbol{s}_1, \ldots, \boldsymbol{s}_k \}$ of ONBs, which may be chosen deterministically or randomly. For each setting $\boldsymbol{s}$, independent measurements are performed on $m$ identical copies of the state, and the counts $N(\boldsymbol{o} \vert \boldsymbol{s})$ are recorded, where $N(\boldsymbol{o} \vert \boldsymbol{s})$ is the number of times the outcome $\boldsymbol{o}$ is observed when measuring in setting $\boldsymbol{s}$. The total number of quantum samples is therefore $N=m \times k$. The resulting dataset $\mathcal{D}:= \{ N(\boldsymbol{o} \vert \boldsymbol{s}) : \boldsymbol{o}\in \{+1, -1 \}^n, \boldsymbol{s} \in \mathscr{S} \} $ is a $2^n \times k$ table whose columns are independent and contain all the counts in a given setting. Its probability  is given by the product of multinomials
\begin{equation}\label{eq1:likelihood}
p_{\rho}(\mathcal{D} \vert \mathcal{S}) = \prod_{\boldsymbol{s}} \frac{m!}{\prod_{\boldsymbol{o}} N(\boldsymbol{o} \vert \boldsymbol{s})!} \prod_{\boldsymbol{o}} p_{\rho}(\boldsymbol{o} \vert \boldsymbol{s})^{N(\boldsymbol{o} \vert \boldsymbol{s})} 
\end{equation}
Our goal is to statistically reconstruct the density matrix $\rho$ from the counts dataset 
$\mathcal{D}$. This can be seen as a statistical inverse problem of reversing the measurement map
\begin{eqnarray*}
\mathcal{M} : M(\mathbb{C}^d)& \to& \mathbb{R}^{d\cdot k}\\
\mathcal{M} : \rho &\mapsto & p_{\rho}(\cdot \vert \mathscr{S}) 
\end{eqnarray*}
in the sense that given a sample $\mathcal{D}$ from $p_{\rho}(\cdot \vert \mathcal{S}) $, one would like to produce an estimator $\hat{\rho} (\mathcal{D})$ which is close to $\rho$ with respect to some meaningful distance measure. The next section lists several figures of merit considered here.


%
%



%

 \subsection{ Error functions}
 
 {\setlength{\extrarowheight}{7pt}
 \begin{table}
\centering
\begin{tabularx}{\textwidth}{|c|X|}
\hline
Error function & Definition \\ 
\hline
Frobenius norm squared & $
\Vert \hat{\rho} - \rho \Vert_2^2 ={\rm Tr}\left[ (\hat{\rho}-\rho)^2 \right]$  \\ \hline 
Trace norm & $
\Vert \hat{\rho} - \rho \Vert_{1}= 
{\rm Tr} |\hat{\rho}-\rho | $ \\ \hline
Operator norm & $
\Vert \rho - \sigma \Vert= \lambda_{\rm max} 
(|\hat{\rho}-\rho|) $ \\ \hline
Bures distance & $D_{B}(\hat{\rho}, \rho)^2 :={ 2 \left( 1- {\rm Tr}\left[\sqrt{\sqrt{\rho} \hat{\rho} \sqrt{\rho}} \right] \right)}$\\ \hline 
Hellinger distance & $D_H(\hat{\boldsymbol{\lambda}}, \boldsymbol{\lambda})^2 := { 2 \left( 1- \sum_i^d \sqrt{\hat{\lambda}_i \lambda_i} \right)}$  \\ \hline
\end{tabularx}
\caption{ The different error functions used to measure the distance between the true state and the estimator. The Bures distance is defined only for states $\hat{\rho}, \rho \in \mathcal{S}_d$, and its classical analogue the Hellinger distance is defined between two probability distributions. }   
\label{tb1:errorfunctions}
\end{table}  
 
Let us denote the \emph{risk} or mean error of an estimator $\hat{\rho}$ as $\mathbb{E} \left[ D(\hat{\rho}, \rho) \right]$, where $D(\hat{\rho}, \rho)$ represents a particular error function. In our theoretical analysis and simulations study we estimate the risk for several choices of the error function $D(\hat{\rho} ,\rho)$, which are tabulated in Table \ref{tb1:errorfunctions}. Note that the Bures distance is defined only over the space of density matrices, and therefore applies only in the case where the estimators are density matrices. The classical analogue of the Bures distance is the Hellinger distance between two probability distributions. Here we consider the  Hellinger distance $D_{H}(\hat{\boldsymbol{\lambda}}, \boldsymbol{\lambda})^2$ between the eigenvalues $\{\lambda_1,\dots, \lambda_d\}$ of the true state and those of the estimator $\{\hat{\lambda}_1, \dots, \hat{\lambda}_d\}$, seen as probability distributions. Later on we will show that the behaviour of the Bures distance error  is strongly correlated with that of the squared Hellinger distance error.

\section{Fisher information, asymptotic normality and maximum likelihood}\label{sec1:Fisher} 

The maximum likelihood (ML) estimator  is one of the most commonly used  and well understood statistical tools with a universal range of applicability. Before considering its use in quantum tomography, we briefly review some of the key concepts and results related to the ML estimator and its asymptotic behaviour \cite{LehmannCasella,YoungSmith}. Consider the scenario in which we are given $N$ independent samples $X_1, \dots , X_N$ from a discrete probability distribution $p_{\boldsymbol{\theta}}$ over a countable set $\Omega$; the probability distribution is assumed to depend \emph{smoothly} on an unknown parameter $\boldsymbol{\theta}$ which belongs to an \emph{open} subset $\Theta$ of $\mathbb{R}^p$. The likelihood of the dataset is 
$p_{\boldsymbol{\theta}}(X_1, \dots , X_N)= \prod_{j=1}^N p_{\boldsymbol{\theta}}(X_j)$. The ML estimator is the point in $\Theta$ where the likelihood of the observed data attains its maximum value
$$
\hat{\boldsymbol{\theta}}_{\rm ML} := \underset{\boldsymbol{\theta}^\prime\in \Theta}{\arg\max} ~p_{\boldsymbol{\theta}^\prime}(X_1, \dots , X_N).
$$
The likelihood can be expressed in terms of the \emph{empirical distribution} of the data 
$\boldsymbol{f} = \sum_j\delta_{X_j}/N$ which collects the frequencies of different outcomes in 
$\Omega$. Indeed the log-likelihood can be written as (cf. \eqref{eq1:likelihood})
\begin{eqnarray*}
\ell_{\boldsymbol{\theta}} &:=& \log p_{\boldsymbol{\theta}}(X_1, \dots , X_N) = 
N\sum_{i\in \Omega}  \boldsymbol{f}(i) \log p_{\boldsymbol{\theta}} (i) + C(\boldsymbol{f})\\
&=& 
- N K( \boldsymbol{f}\| p_{\boldsymbol{\theta}}) + C^\prime(\boldsymbol{f}) 
\end{eqnarray*}
where $K(\cdot\|\cdot)$ relative entropy (Kullback-Leibler divergence), and $C, C^\prime$ do not depend on $\boldsymbol{\theta}$. The ML estimator is thus the closest point to $\boldsymbol{f}$ with respect to the relative entropy 
$$
\hat{\boldsymbol{\theta}}_{\rm ML} = \underset{\boldsymbol{\theta}^\prime\in \Theta}{\arg\min}\,
K (\boldsymbol{f} \, \| \, p_{\boldsymbol{\theta}^\prime} )
$$
\emph{Asymptotics.} The appeal of the ML estimator lies in part in its asymptotic optimality properties. For large enough sample sizes $N$, a central limit behaviour emerges and the ML estimator becomes normally distributed around the true parameter, with a variance shrinking at the rate $1/N$
\begin{equation}\label{eq.MLE}
\sqrt{N} (\hat{\boldsymbol{\theta}}_{\rm ML}  - \boldsymbol{\theta} ) 
\overset{\mathcal{D}}{\longrightarrow} 
N(0, I^{-1} (\boldsymbol{\theta}) ).
\end{equation}
The convergence above holds in distribution as $N\to \infty$ and the limit is a centred Gaussian distribution with covariance given by the inverse of the \emph{Fisher information matrix}. The latter is a $p\times p$ positive matrix 
$$
I (\boldsymbol{\theta})_{i,j} = 
\mathbb{E}_{\boldsymbol{\theta}} 
\left[ 
\frac{\partial \log p_{\boldsymbol{\theta}} (X)}{\partial \boldsymbol{\theta}_i}
\frac{\partial \log p_{\boldsymbol{\theta}}(X)}{\partial \boldsymbol{\theta}_j}
\right]
$$
which encapsulates how informative a single sample $X$ from $p_{\boldsymbol{\theta}}$ is about $\boldsymbol{\theta}$. The limiting covariance $ I^{-1} (\boldsymbol{\theta})$ is equal to the lower bound appearing in the the Cram\'{e}r-Rao inequality for unbiased estimators. Estimators which satisfy equation \eqref{eq.MLE} are called \emph{efficient} and statistical theory shows that one cannot improve on their asymptotic accuracy, except on a measure zero set of the parameter space \cite{LehmannCasella}. In particular if $d(\cdot, \cdot)$ is a locally quadratic (positive) loss function on $\Theta$,
$$
d(\hat{\boldsymbol{\theta}}, \boldsymbol{\theta}) = 
(\hat{\boldsymbol{\theta}}-\boldsymbol{\theta})^T 
G
(\boldsymbol{\theta})
(\hat{\boldsymbol{\theta}}-\boldsymbol{\theta}) + o(\|\hat{\boldsymbol{\theta}}-\boldsymbol{\theta}\|^2 )
$$
then the risk of the MLE satisfies
$$
N \mathbb{E} [d(\hat{\boldsymbol{\theta}}_{\rm ML}, \boldsymbol{\theta}) ] \longrightarrow {\rm Tr} (G(\boldsymbol{\theta}) I^{-1}(\boldsymbol{\theta})).
$$
It is important to note that the \emph{asymptotic normality} property \eqref{eq.MLE} relies on the smoothness of the model and the fact that $\boldsymbol{\theta}$ is an interior point of the parameter space 
$\Theta$. For large $N$ the ML estimator lies (with high probability) within in a small neighbourhood of $\boldsymbol{\theta}$ of size $1/\sqrt{N}$, and the parameter space looks like $\mathbb{R}^p$ for all practical purposes.  However, when $\Theta$ has a boundary, the ML estimator will not be asymptotically normal for parameters  $\boldsymbol{\theta}$ lying on the boundary, and one needs to analyse such models more carefully. This is the case in quantum tomography, when the unknown state is not full-rank and therefore lies on the boundary of 
$\mathcal{S}_d$, or is so close to the boundary that the asymptotic theory will kick in only for sample sizes that are much larger than those obtained in real experiments.
\subsection{The maximum likelihood estimator in quantum tomography}
\label{sec.mle.tomo}
 
We will now discuss in more detail the properties of the MLE in the specific case of quantum tomography. Given the measurement data encoded in the dataset $\mathcal{D}$, the MLE is defined as the maximum of $p_{\tau}(\mathcal{D} \vert \mathscr{S})$
%
over  $\tau\in\mathcal{S}_d$. By passing to log-likelihood and  discarding the constant factorial terms in (\ref{eq1:likelihood}), we obtain the following form of the estimator
\begin{equation}\label{eq.ml.tomo}
\hat{\rho}_{\rm ML} := \underset{\tau\in \mathcal{S}_d}{\arg\max} \sum_{\boldsymbol{o}, \boldsymbol{s}} \boldsymbol{f} (\boldsymbol{o} \vert \boldsymbol{s}) p_{\tau}(\boldsymbol{o} \vert \boldsymbol{s})
= \underset{\tau\in \mathcal{S}_d}{\arg\min} \sum_{\boldsymbol{s}} 
K( \boldsymbol{f} (\cdot \vert \boldsymbol{s}) \| p_{\tau}(\cdot \vert \boldsymbol{s}))
\end{equation}
where $\boldsymbol{f} (\boldsymbol{o} \vert \boldsymbol{s}) = N (\boldsymbol{o} \vert \boldsymbol{s})/m$ are the empirical frequencies of the data.
The MLE is commonly used in quantum tomography \cite{MLparis,MLrobin,MLmofqubits,LocalML}, and several implementation methods have been proposed including Hradil's iterative algorithm \cite{HradilML}. Our specific implementation uses the CVX package for disciplined convex programming on MATLAB \cite{cvx}. The general asymptotic theory guarantees that the ML estimator is asymptotically normal for \emph{full rank states}, i.e. the interior of 
$\mathcal{S}_d$. To get more insight into the general asymptotic behaviour for a given state 
$\rho$, we will choose a \emph{local} parametrisation defined in terms of the matrix elements of $\rho$ with respect to its eigenbasis. Let $\lambda_1\geq \lambda_2 \dots, \geq \lambda_d>0$ be the eigenvalues of a full-rank state $\rho$. Then any neighbouring state can be written as 
$$
\rho^\prime =\rho_{\boldsymbol{\theta}} = 
\rho+ \delta\rho_{\boldsymbol\theta}
$$
where $\delta\rho_{\boldsymbol\theta}$ is trace zero matrix which is completely  parametrised by
\begin{align}
\boldsymbol{\theta} &:= 
\left(  \theta^{(r)} ;\theta^{(i)} ; \theta^{(d)}\right) 
\label{eq1:parametrisation} \\
&= (
{\rm Re}\delta\rho^\prime_{1,2}, \ldots, {\rm Re}\delta\rho^\prime_{d-1,d}\, ; \, 
{\rm Im}\delta\rho^\prime_{1,2}, \ldots, {\rm Im}\delta\rho^\prime_{d-1,d}\, ; \, 
\delta\rho^\prime_{2,2}, \ldots, \delta\rho^\prime_{d,d} 
) \in \mathbb{R}^{d^2-1} \notag.
\end{align}
The ML estimator $\hat{\rho}_{\rm ML}=\rho_{\hat{\boldsymbol{\theta}}_{\rm ML}} $ has parameter $\hat{\boldsymbol{\theta}}_{\rm ML}$ which is normally distributed around $0$ with covariance $I^{-1}(\rho | \mathscr{S})/N$ where the Fisher information is the average of the individual informations for different sets  $\mathbf{s}\in \mathscr{S}$  
\begin{equation}\label{eq1:Fisher}
I(\rho \vert \mathscr{S}) = \frac{1}{k} \sum_{\mathbf{s} \in \mathscr{S}} I(\rho \vert \mathbf{s}), \qquad
I(\rho \vert \mathbf{s})_{a,b} :=  \sum_{\mathbf{o}: p(\mathbf{o}\vert \mathbf{s}) >0} \frac{1}{p_{\rho}(\mathbf{o}\vert\mathbf{s})}\frac{\partial p_{\rho}(\mathbf{o}\vert \mathbf{s})}{\partial \theta_{a}} \cdot \frac{\partial p_{\rho}(\mathbf{o}\vert \mathbf{s})}{\partial \theta_{b}}.
\end{equation}
%
In particular, for any locally quadratic loss function $d(\cdot, \cdot)$ 
(e.g. Frobenius distance, or Bures distance) with weight matrix $G(\boldsymbol{\theta})$, 
the associated risk has the asymptotic behaviour
\begin{equation}\label{eq1:MSE}
N\mathbb{E}\left[d(\rho_{\boldsymbol{\theta}}, \hat{\rho}_{\rm ML}) \right] \longrightarrow  {\rm Tr}\left( I(\rho \vert \mathscr{S})^{-1}G(\boldsymbol{\theta})\right).
\end{equation} 

Now, let us assume that the unknown state $\rho$ is on the boundary of $\mathcal{S}_d$, and more precisely belongs to the space  of rank $r$ states $\mathcal{S}_d(r)\subset M(\mathbb{C}^d)$, for a \emph{fixed and known} rank $r\leq d$. In its own eigenbasis, $\rho$ is the diagonal matrix of eigenvalues ${\rm Diag}(\lambda_1, \dots, \lambda_r, 0,\dots, 0)$, and any sufficiently close state is uniquely determined by its matrix elements in the first $r$ rows (or columns). Intuitively this can be understood by noting that any rank-$r$ state $\rho^\prime$ in the neighbourhood of $\rho$ can be obtained by perturbing the eigenvalues and performing a small rotation of the eigenbasis; in the first order of approximations these transformation leave the $(d-r)\times (d-r)$ lower-right corner unchanged so 
\begin{equation}\label{eq.rho.eigenbasis}
\rho^{\prime} =
\left(
\begin{array}{cc}
D & 0\\
0 &0
\end{array}
\right)
+ 
 \left(
\begin{array}{cc}
A
& 
B
\\
B^*
& 
C
\end{array}
\right), \qquad D:= {\rm Diag}(\lambda_1, \dots, \lambda_r)
\end{equation}
where $C = O(\|A\|^2, \|B\|^2)$. We therefore choose the (local) parametrisation $\rho^{\prime} = \rho_{\boldsymbol{\theta}}$ 
where only the matrix elements of $A$ and $B$ enter the parameter 
$\boldsymbol{\theta}$. For this model, any rank-$r$ state corresponds to an interior point of the parameter space $\mathcal{S}_d(r)$, and consequently the ML estimator obeys the asymptotic normality theory described above. 

However, if the rank of $\rho$ is not known in advance, then the diagonal block $C$ of 
$\rho^\prime$ needs to be included in the model in order to describe  neighbouring states of higher rank. As $\rho^\prime$ is a state, the block $C$ is a positive matrix, and therefore its matrix elements are constrained. This complicates the analysis of the likelihood function, and invalidates the asymptotic normality of the ML estimator.

What should be the theory replacing asymptotic normality ? At the moment there isn't a complete answer to this question, but some important progress has been made in \cite{BlumeKohout_2018}. Following this work and \cite{spectralthresholding}, it is instructive to study an extended, `non-physical' model in which the positivity requirement is dropped and (locally around $\rho$) the parameter space is taken to be that of selfadjoint matrices of trace-one $M_{sa}^1(\mathbb{C}^d)$. Therefore, the `unphysical' parameter $\boldsymbol{\theta}$ consists of the matrix elements of the blocks $A,C,B$ (except one diagonal element due to normalisation), with $\rho^\prime=\rho_{\boldsymbol{\theta} }$ given by equation \eqref{eq.rho.eigenbasis}.
We now make the `mild' assumption that $p_{\rho}(\boldsymbol{o}|\boldsymbol{s})>0$ for all pairs $(\boldsymbol{o},\boldsymbol{s})$; indeed this condition is satisfied for `generic' states but fails for states whose support is orthogonal to some of the measurement basis vectors. Under this assumption, we find that locally around $\rho$, we can define a statistical model given by   bona-fide probability distributions
$
p_{\boldsymbol{\theta} } := p_{\rho_{\boldsymbol{\theta} }}.
$
We can therefore define the \emph{unconstrained} maximum likelihood (uML) estimator as in equation \eqref{eq.ml.tomo}
where the optimisation is performed over the unconstrained local neighbourhood 
$\mathcal{O}\subset M^1_{sa}(\mathbb{C}^d)$ of $\rho$, 
rather than over the space of states $\mathcal{S}_d$. Since $\rho$ is on the boundary of $\mathcal{S}_d$, the probability that 
$\hat{\rho}_{\rm uML}$ falls outside the state space is significant and does not vanish in the asymptotic limit. In this case, $\hat{\rho}_{\rm uML}$ does not coincide with the `constrained' or regular ML estimator $\hat{\rho}_{\rm ML}$. In fact, each of them can be seen as the projection with respect to the relative entropy distance of the empirical distribution $\boldsymbol{f}$  onto the sets of probabilities $\mathcal{M}(\mathcal{O})$ and respectively $\mathcal{M}(\mathcal{S}_d)$,  
 where $\mathcal{M}:M(\mathbb{C}^d)\to \mathbb{R}^{k\cdot d}$ is the measurement map.

\emph{Asymptotics.} 
Since $\rho$ corresponds to an interior point of the \emph{unconstrained} parameter space, the general asymptotic normality theory applies again. In particular, the uML estimator is normally distributed around 
$\boldsymbol{\theta}=0$ with variance $(N I(\rho | \mathscr{S}))^{-1}$, where $I(\rho | \mathscr{S})$ is the Fisher information \eqref{eq1:Fisher} computed with respect to the unconstrained parametrisation. Moreover
%
by Taylor expanding the log-likelihood function around the $\boldsymbol{\theta}_{\rm uML}$, and using the fact the the first derivative vanishes at this point, we obtain the second order approximation 
\begin{equation}\label{eq.likelihood}
 \ell(\boldsymbol{\theta})-  \ell(\boldsymbol{\theta}_{\rm uML})  \approx
- 
\frac{N}{2}( \boldsymbol{\theta}- \hat{\boldsymbol{\theta}}_{\rm uML} )^T 
I(\rho | \mathscr{S}) 
( \boldsymbol{\theta}- \hat{\boldsymbol{\theta}}_{\rm uML} ). 
\end{equation}
This implies tht for large $N$, the ML estimator $\hat{\rho}_{\rm ML}$ is the projection of 
$\hat{\rho}_{\rm uML}$ onto $\mathcal{S}_d$ with respect to the quadratic distance determined by the Fisher information
\begin{equation}\label{eq.fisher.distance}
d_I (\rho_{\boldsymbol{\theta}}, \rho_{\boldsymbol{\theta}^\prime}):= (\boldsymbol{\theta} - \boldsymbol{\theta}^\prime)^T 
I(\rho | \mathscr{S}) 
(\boldsymbol{\theta} - \boldsymbol{\theta}^\prime).
\end{equation}
Moreover, the probability distribution of the MLE is obtained by projecting the Gaussian distribution corresponding to $\hat{\rho}_{\rm uML}$, onto $\mathcal{S}_d$. Although the projection can be computed efficiently using convex optimisation, finding a general characterisation of the resulting distribution seems to be a challenging problem \cite{SuessGross}. Nevertheless, \cite{BlumeKohout_2018} shows that the problem is tractable in those cases where the metric $d_I$ is (approximately) isotropic, so that random matrix results such as the emergence of Wigner semicircle law can be used to treat the asymptotic theory. In section \ref{sec.PLS.covariant} we will use these ideas to study the asymptotic behaviour of the projected least squares estimator.

\section{The least squares estimator}\label{sec.LS}

The least squares (LS) estimator \cite{QiBo,spectralthresholding} is based on the observation that the outcome probabilities $p_\rho (\boldsymbol{o} | \boldsymbol{s})$ depend linearly on the state $\rho$. Let 
\begin{equation}\label{eq1:paraLS}
\rho= \sum_{i=1}^{d^2} \beta_i  \tau_i
\end{equation}
be the decomposition with respect to an orthonormal basis of $M(\mathbb{C}^d)$ consisting of selfadjoint matrices $\{\tau_i: 1\leq i\leq d^2\}$, and 
$\boldsymbol{\beta}:= (\beta_1, \dots , \beta_{d^2})^T$ the corresponding vectorisation. 
Then the probabilities can be expressed as 
$$
p_\rho (\boldsymbol{o} | \boldsymbol{s}) = {\rm Tr} (\rho P_{\boldsymbol{o}}^{\boldsymbol{s}})=
\sum_i X_{(\boldsymbol{o}|\boldsymbol{s})}^i \beta_i, \qquad 
X_{(\boldsymbol{o}|\boldsymbol{s})}^i := {\rm Tr} (\tau_i P_{\boldsymbol{o}}^{\boldsymbol{s}} )
$$
where $X$ is a $kd\times d^2$ fixed matrix depending on the measurement and the chosen vectorisation. In an experimental setup we do not have access to the true probability vector. 
Instead from the $d \times k$ dataset of counts, we can compute the empirical probabilities $f(\boldsymbol{o} \vert \boldsymbol{s}) := N(\boldsymbol{o} \vert \boldsymbol{s})/m$, whose expectations are $\mathbb{E} f(\boldsymbol{o} \vert \boldsymbol{s}) = p_\rho(\boldsymbol{o} \vert \boldsymbol{s})$. 
Replacing probabilities vector with empirical frequencies 
\begin{equation}\label{eq:matrixform}
\boldsymbol{f} = 
X \boldsymbol{\beta} + \boldsymbol{\epsilon} , 
\end{equation}
where $\boldsymbol{\epsilon} \in \mathbb{R}^{dk}$ is a mean zero vector of statistical noise.
The noise distribution is irrelevant for the definition of the LS estimator, but we will return to this when discussing its error, and the generalised least squares estimator.

The least squares solution to the above system of equations is defined as the minimiser of the following optimisation problem 
\begin{equation}\label{eq1:LSmin}
\hat{\boldsymbol{\beta}}_{\rm LS}:= \underset{v \in \mathbb{R}^{d^2}}{\arg\min} \Vert X v - \boldsymbol{f}  \Vert^2
\end{equation}
and has the explicit form $\hat{\boldsymbol{\beta}}_{\rm LS} = (X^TX)^{-1} \cdot X^T \cdot \boldsymbol{f}$. The final estimate $\hat{\rho}_{\rm LS}$ of the density matrix  is then constructed from the estimated parameter vector $\hat{\boldsymbol{\beta}}_{\rm LS}$ using equation \eqref{eq1:paraLS}. We note that the least squares estimator produces a  state estimate 
$\hat{\rho}_{\rm LS} = \rho_{\boldsymbol{\beta}_{\rm LS}}$ that is not necessarily a density matrix. 

\emph{Least squares as a projection.} Let $\mathcal{M}:M(\mathbb{C}^d)\to \mathbb{C}^{k\cdot d}$ be the map associated to the measurement, and let $\hat{p}_{\rm LS}:=\mathcal{M}(\hat{\rho}_{\rm LS})=  X \hat{\boldsymbol{\beta}}_{\rm LS}$ be the `probability distribution' corresponding to the LS estimator. This belongs to the hyperplane 
$\mathcal{L}_d = \mathcal{M}(M_{sa}^1(\mathbb{C}^d))$ which contains the convex set of probabilities $\mathcal{P}_d = \mathcal{M}(\mathcal{S}_d)$.  The LS estimator \eqref{eq1:LSmin} can then be interpreted as the projection of the frequency vector $\boldsymbol{f}$ onto $\mathcal{L}_d$ with respect to the Euclidian distance in $\mathbb{R}^{k\cdot d}$.

\subsection{Least squares as inversion of a measure-and-prepare channel}\label{sec.ls.inverse}

The LS estimator was introduced above by choosing a particular vectorisation of the density matrix. While this is useful for numerical implementations, the disadvantage is that one loses the physical interpretation of the quantum state and the measurement map. Conceptually, it is useful to define the LS estimator in a ``coordinate free" way which can be interpreted as the inversion of a certain measure-and-prepare map associated to the measurement \cite{GutaKahnKungTropp}. 
Let $\mathcal{D}:M(\mathbb{C}^d)\to M(\mathbb{C}^d)$ be the quantum channel
$$
\mathcal{D}: \rho\mapsto \frac{1}{|\mathscr{S}|} \sum_{\boldsymbol{o}, \boldsymbol{s}} 
{\rm Tr} (\rho P_{\boldsymbol{o}}^{\boldsymbol{s}}) \cdot P_{\boldsymbol{o}}^{\boldsymbol{s}}
$$
which is the composition $\mathcal{P}\circ\mathcal{M}$ of the measurement with collection of bases $\mathscr{S}$, and the preparation map where the pure state $P_{\boldsymbol{o}}^{\boldsymbol{s}}$ is prepared for each outcome $\boldsymbol{o}$ of the measurement in basis $\boldsymbol{s}$. If $\rho$ is represented in its vectorised form, then the map $\mathcal{D}$ is given by the matrix $X^TX/|\mathscr{S}|$. On the other hand, the preparation map $\mathcal{P}$ has matrix $X^T$, so that the LS estimator can be expressed as 
\begin{equation}\label{eq.ls.channel}
\hat{\rho}_{\rm LS} = \mathcal{D}^{-1} \circ \mathcal{P} (\boldsymbol{f}/|\mathscr{S}|).
\end{equation}
From this expression we see immediately that ${\rm Tr}(\hat{\rho}_{\rm LS})=1$ as consequence of the fact that $\boldsymbol{f}/|\mathscr{S}|$ is a probability distribution and $\mathcal{D}$ is trace preserving. Additionally, we note the the accuracy of the LS estimator is linked to the properties of the channel $\mathcal{D}$. 
In particular, in the case of Pauli measurements the channel is given 
by a tensor product $\mathcal{D} = \mathcal{C}^{\otimes n}$ of qubit depolarising channels \cite{spectralthresholding,GutaKahnKungTropp}
$$
\mathcal{C}: \rho\mapsto \frac{1}{3} \rho + \frac{2}{3}\frac{ \mathbf{1}}{2}.
$$
On the other hand, the measurements corresponding to the class of 2-designs (which includes covariant measurements, mutually unbiased bases, stabiliser states, symmetric informationally complete measurements) have associated channel given by \cite{GutaKahnKungTropp}
$$
\mathcal{D}:\rho \mapsto \frac{1}{d+1} \rho + \frac{d}{d+1} \frac{\mathbf{1}}{d}.
$$

\subsection{Concentration bounds and asymptotic behaviour of the LS}

For more insight into the structure of the LS estimator let us unpack equation \eqref{eq.ls.channel} and note that $\hat\rho_{\rm LS} $ can be written as an average of independent, identically distributed matrices 
$A^{(1)},\dots ,A^{(m)}$ with $\mathbb{E}(A^{(i)}) =\rho$
\begin{equation}\label{eq.LS.average}
\hat\rho_{\rm LS} 
=   \frac{1}{m}\sum_{i=1}^m A_i
= \frac{1}{m}\sum_{i=1}^m \frac{1}{|\mathscr{S}|} \sum_{\bf s\in \mathscr{S}} \mathcal{D}^{-1} ( P^{\bf s}_{{\bf o}_{i| {\bf s}}}) 
\end{equation}
where ${\bf o}_{i| {\bf s}}$ are the outcomes of the measurements with respect to setting ${\bf s}$. 

In this form, the LS estimator is amenable to non-asymptotic matrix concentration techniques, as well as asymptotic normality theory. The following result \cite{GutaKahnKungTropp} was obtained by applying matrix Bernstein inequalities \cite{tropp_user-friendly_2012} and provides a non-asymptotic confidence bound for LS with respect to the operator norm distance, see also \cite{spectralthresholding}.
\begin{theorem}\label{th.LS}
Let $\hat\rho_{\rm LS}$ be the LS estimator of $\rho$ for a dataset consisting of $N=m \times k$ samples. 
Then 
\begin{equation*}
\mathrm{Pr} \left[ \left\| \hat{\rho}_{\rm LS} - \rho \right\| \geq \epsilon \right] \leq d 
\mathrm{e}^{ - \frac{3N \epsilon^2}{8g(d)} } \quad \epsilon \in \left[0,1 \right].
\end{equation*}
where $g(d)  =2 d $ for $2-$design measurements and $g(d) \simeq  d^{1.6}$ for Pauli measurements.
\end{theorem}
The theorem provides upper bounds for risks with respect to commonly used loss functions such as the Frobenius distance (norm-two squared) and the trace-norm distance, see also \cite{SugiyamaTurnerMurao} for related results. Indeed using the basic inequalities 
$\|A\|_2^2\leq d \| A\|^2 $ and $\|A\|_1\leq d \| A\|$ we obtain the upper bounds 
\begin{equation}\label{eq.risk.ls.2design}
\mathbb{E} \|\rho - \hat{\rho}_{\rm LS}\|_2^2 \leq c_2\log(d) \frac{d^2}{N},\qquad
\mathbb{E} \|\rho - \hat{\rho}_{\rm LS}\|_1 \leq c_1\log(d) \frac{d \sqrt{d}}{\sqrt{N}}
\end{equation}
for the 2-design measurements and 
\begin{equation}\label{eq.risk.ls.Pauli}
\mathbb{E} \|\rho - \hat{\rho}_{\rm LS}\|_2^2 \leq C_2\log(d) \frac{d^{2.6}}{N},\qquad
\mathbb{E} \|\rho - \hat{\rho}_{\rm LS}\|_1 \leq C_1\log(d) \frac{d^{1.8}}{\sqrt{N}}
\end{equation}
for the Pauli basis measurement, where $c_1, c_2, C_1, C_2$ are a numerical constants. Moreover, in \cite{GutaKahnKungTropp} it was shown that the log factor in \eqref{eq.risk.ls.2design} can be removed when we deal with covariant measurements. On the other hand, in section \ref{sec1:Fisher} we have shown that for the maximally mixed state, the optimal convergence rate for the Frobenius distance is of the order $O(d^2/N)$; this means that the upper bound \eqref{eq.risk.ls.2design} cannot be improved, when seen as a uniform bound over \emph{all} states. However, this leaves open the possibility that special classes of states can be estimated with higher accuracy that that provided by the LS estimator. Indeed, we will see that simple modifications of the LS estimator which take into account the positivity of the unknown state can significantly reduce the estimation errors for low rank states.


\emph{Asymptotics.} Thanks to its simplicity, the LS estimator has a tractable asymptotic theory. 
As in the case of the concentration Theorem \ref{th.LS} the key point is that the error
$\rho_{\rm LS} $ is a sum of independent, identically distributed matrices given by equation \eqref{eq.LS.average}.
In the limit of large $m$ each matrix element of $\rho_{\rm LS}$ has a Gaussian distribution; in terms of the vectorised form we have the Central Limit for $m\to \infty$
$$
\sqrt{m} \left[\hat{\boldsymbol{\beta}}_{\rm LS} -\boldsymbol{\beta} \right] \longrightarrow N(0, V_{\rm LS}), \quad V_{\rm LS} = (X^TX)^{-1}  X^T \Omega X (X^TX)^{-1}
$$
where $ \Omega:= m {\rm Cov}\left[ \boldsymbol{\epsilon} \vert \mathscr{S} \right] $ is the (renormalised) covariance of the noise $\boldsymbol{\epsilon}$. Due to the independent structure of the measurement settings, the latter has a non-trivial block diagonal form with $d \times d$ blocks corresponding to the different settings 
$\boldsymbol{s}$ 
\begin{equation}\label{eq1:covariance}
\left[\Omega_{{\bf s}}\right]_{ij}= 
{\rm Cov}\left[ \epsilon_{\boldsymbol{s}} \right]_{ij}= 
p_\rho (\boldsymbol{o}_i \vert \boldsymbol{s}) \delta_{ij} - p_\rho (\boldsymbol{o}_i \vert \boldsymbol{s})p_\rho (\boldsymbol{o}_j \vert \boldsymbol{s}).
\end{equation}
This allows us in principle to compute the asymptotic risk of an arbitrary quadratic loss function with weight matrix $G$ (such as the Frobenius distance) as
$$
m\mathbb{E} \left[ (\hat{\boldsymbol{\beta}}_{\rm LS} - \boldsymbol{\beta} )^T G   (\hat{\boldsymbol{\beta}}_{\rm LS} - \boldsymbol{\beta} )\right]
\longrightarrow
{\rm Tr}(G V_{\rm LS}).
$$
Since $\boldsymbol{f}$ is a vector of frequencies, it must satisfy $k$ normalisation constraints, which is reflected in the fact each block $\Omega_{\bf s}$ is singular, with zero eigenvector 
$\mathbf{1} = (1,\dots, 1)^T$. Consequently, the covariance matrix $V_{\rm LS}$ is singular with the zero eigenvector corresponding to the trace which is a fixed parameter. We will come back to this when discussing the generalised least squares estimator.

\subsection{Asymptotic theory of LS for covariant measurements}\label{sec.asymptotics.ls.covariant}

While asymptotic normality tells us that the estimator $\hat{\boldsymbol{\beta}}_{\rm LS}$ lies in 
an ellipsoid centred at $\boldsymbol{\beta}$, it treats the estimator as a vector rather that as a matrix. For instance, it does not immediately provide an asymptotic theory for the operator norm error $\|\hat{\rho}_{\rm LS} - \rho\|$. Random matrix theory provides us with other asymptotic results which take into account the matrix structure of the estimator. To explore this avenue we will make a simplifying assumption and place ourselves in the scenario of covariant measurements. The outcome of such a measurement is a one dimensional projection $P= |\psi \rangle \langle \psi|$ and the corresponding positive operator valued measure is
\begin{equation}\label{eq.povm.covariant}
M({\rm d} P) : = d\cdot P \cdot  {\rm d} P 
\end{equation}
where $ {\rm d}P$ is the uniform measure over the space of one dimensional projections; the latter is the measure induced by the Haar measure  over the unitaries $U$ via the action $ P \mapsto UPU^*$. 
An alternative way of obtaining a measurement sample is to choose a random basis and perform a measurement with respect to the chosen basis.

In this setup, the channel $\mathcal{D}$ is given by \cite{GutaKahnKungTropp}
$$
\mathcal{D} :\rho \mapsto 
\int {\rm Tr} (\rho M({\rm d}P)) P =
\int d P   {\rm Tr} (\rho P) {\rm d} P  = 
\frac{1}{d+1} (\rho + \mathbf{1})
$$
and the LS estimator given by equation \eqref{eq.LS.average}  can be written as 
$$
\hat\rho_{\rm LS} = 
\frac{d+1}{N} \sum_{i=1}^N \left( P_i - \frac{\mathbf{1}}{d+1}\right)
$$
where $P_i$ are the independent outcomes of measurements.

We will be particularly interested in the behaviour of the LS estimator for low rank states, as well as states close to the maximally mixed one. Due to covariance it suffices to choose states which are diagonal in the standard basis $\{|i\rangle: i=1, \dots , d\}$, and for simplicity we will restrict ourselves to rank-r states with equal eigenvalues $\rho_r = \sum_{i=1}^r |i\rangle\langle i| /r$. 
As in section \ref{sec.mle.tomo} we write the LS estimator as
\begin{equation}\label{eq.rho.ls.blocks}
\hat{\rho}_{\rm LS}  =
\frac{1}{r}\left(
\begin{array}{cc}
I_r & 0\\
0 &0
\end{array}
\right)
+ \frac{1}{\sqrt{N}}
 \left(
\begin{array}{cc}
A
& 
B
\\
B^*
& 
C
\end{array}
\right), \qquad 
\end{equation}
where $A,B,C$ are block matrices of mean zero, and $I_r$ is the $r\times r$ identity block. 
We will deal with each block separately.

{\it Block C}. By covariance, the distribution of the block $C$ is invariant under unitary transformations in 
$\mathbb{C}^{d-r}$. As the LS estimator is unbiased, the matrix elements of $C$ 
are centred. The off-diagonal elements have real and imaginary parts with variances equal to
$$
 {\rm Var}({\rm Re}C_{ij}) = {\rm Var}({\rm Im}C_{ij})= \frac{1}{2}\mathbb{E} |C_{ij}|^2 =\frac{v_c}{2}
 $$ 
 with 
\begin{eqnarray*}
v_c &=&  
d(d+1)^2 \int \langle U k|\rho_r | U k\rangle \cdot |\langle i|U|k\rangle|^2 \cdot 
 | \langle k | U^* | j\rangle|^2 \,{\rm d}U\\
&=&  d(d+1)^2\int U_{1k} U_{ik} U_{jk}  U^*_{k1} U^*_{ki} U^*_{kj} \, {\rm d}U\\
&=& \frac{d(d+1)^2 }{d (d+1) (d+2 )} = \frac{d+1 }{d+2 }
\end{eqnarray*}
where we have written $P= U |k\rangle\langle k| U^*$ for $r<k \neq i,j$ and replaced the integration over projections with that over unitaries; the integral was then evaluated using Weingarten formulas \cite{Collins2006}. Similarly, the variance and covariance of the diagonal elements are 
$$
{\rm Var }(C_{ii} ) = \frac{d}{d+2}
\qquad 
{\rm Cov }(C_{ii}, C_{jj} ) =- \frac{1}{d+2}
$$ 
and all off-diagonal elements (of all blocks $A,B,C$) have zero covariance with other matrix elements. By the Central Limit Theorem, in the limit of large $N$, the off-diagonal elements of 
$\hat{\rho}_{\rm LS}$ become normally distributed and independent of all other elements. On the other hand, as the covariance of diagonal elements is non-zero, they converge to correlated Gaussian variables. However, if we also take the limit of large $d$ we find that the covariance of the diagonal elements vanishes and the matrix $C$ is distributed approximately as the Gaussian unitary ensemble (GUE). Universality results for random matrices \cite{SchenkerSchulz-Baldes_dependent_2005} imply that the empirical distribution of the eigenvalues of $C/\sqrt{d-r}$ converges weakly to  Wigner's semicircle law  on the interval $[-2,2]$, whose probability density is 
\begin{equation}\label{eq.wigner}
w(x) = \frac{1}{2\pi } \sqrt{4-x^2} . 
\end{equation}
Indeed, panel a) of Figure \ref{fig.wigner} shows a good match between the histogram of the eigenvalues of the error block $C/\sqrt{N}$ for a rank-one state of $n=7$ atoms and $N=10^6$ samples, and the corresponding Wigner distribution.

\begin{figure}[h]
\begin{center}
\includegraphics[width=.45\linewidth]{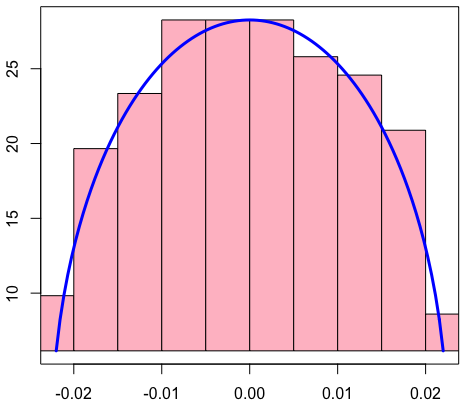}\hspace{8mm}
\includegraphics[width=.45\linewidth]{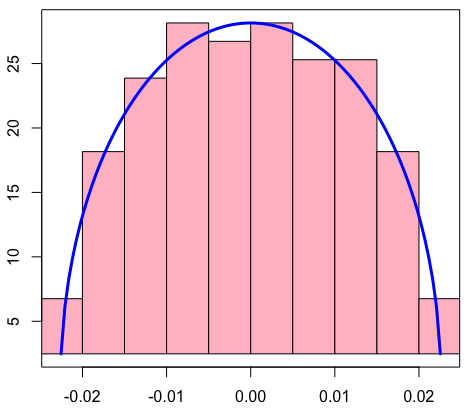}
\caption{Left panel: histogram of eigenvalues of the `error block' $C/\sqrt{N}$ versus the corresponding Wigner distribution, for a $n=7$ atoms, rank $r=1$ state and $N=10^6$ samples. 
Right: histogram of eigenvalues of error matrix for a rank $r=128$ state.}
\label{fig.wigner}
\end{center}
\end{figure}

{\it Block A.} Let us consider now block $A$ of equation \eqref{eq.rho.ls.blocks}. For the same symmetry reasons, its off-diagonal elements (when $r>1$) have real and imaginary parts with variance
$$
{\rm Var}({\rm Re}A_{ij}) = {\rm Var}({\rm Im}A_{ij})= \frac{1}{2}\mathbb{E} |A_{ij}|^2 =\frac{v_a}{2}
$$
with
\begin{eqnarray*}
v_a & = & d(d+1)^2 \int 
\langle U k |\rho_r | U k\rangle \cdot |\langle i | U | k\rangle|^2 \cdot  
| \langle k | U^* | j\rangle|^2 \,{\rm d}U\\
&=&\frac{2d(d+1)^2}{r} \int  U_{ik} U_{ik} U_{jk} U^*_{ki} U^*_{ki} U^*_{kj} \,{\rm d}U +
\frac{(r-2)d(d+1)^2}{r} \int  U_{lk} U_{ik} U_{jk} U^*_{ki} U^*_{kl} U^*_{kj} \,{\rm d}U \\
&=&
\frac{2d(d+1)^2}{r}\cdot \frac{2}{d(d+1)(d+2)} +\frac{(r-2)d(d+1)^2}{r}\cdot \frac{1}{d(d+1)(d+2)} =
\frac{r+2}{r} \cdot \frac{d+1}{d+2}
\end{eqnarray*}
where all indices $i,j,k,l$ are different. 
The diagonal elements have variance
$$
{\rm Var}(A_{ii}) =  
2\frac{d+1}{d+2}\frac{r+2}{r} -\left( 1+ \frac{1}{r}\right)^2 
$$
while the covariance of the diagonal elements is
$$
{\rm Cov}(A_{ii}, A_{jj}) =  
\frac{d+1}{d+2} \frac{r+2}{r} -\left( 1+ \frac{1}{r}\right)^2.
$$
We note that for small $r$ the diagonal elements do not become independent when $N$ and $d$ are large. However, if we also take $r$ to be large, the covariance vanishes, as in the case of the 
$C$ block. By the same argument, the distribution of the eigenvalues of $A/\sqrt{r}$ converges to the Wigner semicircle law \eqref{eq.wigner}. This is illustrated in the right panel of Figure \ref{fig.wigner}, for the case of the totally mixed state of $n=7$ atoms.

{\it Block B.} The elements of block B have variances equal to
$$
{\rm Var}({\rm Re}B_{ij}) = {\rm Var}({\rm Im}B_{ij})= \frac{1}{2}\mathbb{E} |B_{ij}|^2 =\frac{v_b}{2}
$$
with $i\leq r<j$ and 
\begin{eqnarray*}
v_b & = & d(d+1)^2 \int 
\langle U k |\rho_r | U k\rangle \cdot |\langle i | U | k\rangle|^2 \cdot  
| \langle k | U^* | j\rangle|^2 \,{\rm d}U\\
&=&\frac{d(d+1)^2}{r} \int  U_{ik} U_{ik} U_{jk} U^*_{ki} U^*_{ki} U^*_{kj} \,{\rm d}U +
\frac{(r-1)d(d+1)^2}{r} \int  U_{lk} U_{ik} U_{jk} U^*_{k1} U^*_{kl} U^*_{kj} \,{\rm d}U \\
&=&
\frac{d(d+1)^2}{r}\cdot \frac{2}{d(d+1)(d+2)} +\frac{(r-1)d(d+1)^2}{r}\cdot \frac{1}{d(d+1)(d+2)} =
\frac{r+1}{r} \cdot \frac{d+1}{d+2}
\end{eqnarray*}

{\it Eigenvalues distribution of the LS estimator.} 
To better understand the LS estimator, it is instructive to study the distribution of its eigenvalues. This will also be used in the study of the projected least squares estimator furher on. Assuming $N$ is large, we note that $\hat{\rho}_{\rm LS}$ can be written as
\begin{equation}\label{eq.ls.rotation}
\hat{\rho}_{\rm LS} =   U\left( \frac{I_r}{r}  + \frac{D}{\sqrt{N}}\right) U^* + o(N^{-1/2})
\end{equation}
where 
$$
U= \exp\left( \frac{i r\tilde{B}}{\sqrt{N}}\right), \quad
\tilde{B}:= \left(
\begin{array}{cc}
0
& 
iB
\\
-iB^*
& 
0
\end{array}
\right), \quad
D:=
\left(
\begin{array}{cc}
A
& 
0
\\
0
& 
C
\end{array}
\right)
$$
This means that in the leading order in $N^{-1/2}$ the set of eigenvalues of $\hat{\rho}_{\rm LS}$ is the union of those of $I_r/r +A/\sqrt{N}$ and $C/\sqrt{N}$. In the case of large $N,d$ and small rank $r$, the first group of eigenvalues are small fluctuations of size $r/\sqrt{N}$ around $1/r$; in particular these eigenvalues are positive with high probability; the  second group of eigenvalues are distributed according to the Wigner law, and have maximum absolute value approximately equal to $2\sqrt{(d-r)/N}$, and roughly half of them will be negative.

We can now evaluate the asymptotic risk of the LS estimator with respect to the chosen loss functions. 

{\it Frobenius risk.} For the Frobenius distance, the asymptotic mean square error is
\begin{eqnarray*}
N\mathbb{E} \| \hat{\rho}_{\rm LS} - \rho_r\|_2^2 
&=& \mathbb{E}\|A\|_2^2 + 2 \mathbb{E}\|B\|_2^2 + \mathbb{E}\|C\|_2^2\\
&=& \sum_{i=1}^r {\rm Var}(A_{ii}) +2 \sum_{i<j\leq r}  \mathbb{E} |A_{ij}|^2 +
2\sum_{i\leq r <j} \mathbb{E} |B_{ij}|^2 \\
&+&    
\sum_{r<i<j} 2 \mathbb{E} |C_{ij}|^2 + \sum_{i=r+1}^d {\rm Var}(C_{ii})\\
&=& (r+1)(r+2)\frac{d+1}{d+2} - \frac{(r+1)^2}{r}  + 2 \frac{(r+1)(d-r)(d+1)}{d+2} \\
&+& 
\frac{(d-r)(d-r-1)(d+1)}{d+2}  + \frac{(d-r)d}{d+2} .\\
\end{eqnarray*}
In particular, the leading contribution to the Frobenius risk is $d^2/N$, and the dependence with 
$r$ is weak. As illustrated in panels a) and b) of Figure \ref{fig.asymptotics.LS.PLS} the theoretical prediction match the simulation results for 5 and 6 atom states.



{\it Operator norm risk.} The operator-norm error of the LS estimator is 
$$
\|\hat{\rho}_{\rm LS} - \rho_r\| = \frac{1}{\sqrt{N}} 
\left\| \left(
\begin{array}{cc}
A
&
B
\\
B^*
&
C
\end{array}
\right)\right\|
$$
Above we found that the entries of $A, B,C$ become independent in the limit of large $N$, except for correlations between the diagonal elements which vanish if we additionally take the limit of large $d$. Moreover, the variances of the elements in the $B$ block and the off-diagonal elements of $A$ differ from those of the off-diagonal elements of $C$ by factors $(r+2)/r$ and respectively $(r+1)/r$. Therefore, the error block matrix as a whole is not distributed according to GUE ensemble. However, the universality of the Wigner semicircle law, the limit holds not only for highly symmetric ensembles like GUE but also for `small' perturbations of this ensemble, e.g.  random matrices with independent entries, whose variances do not deviate too much from a fixed value \cite{Erdos}.  In particular for low rank $r\ll d$, the total size of the blocks $A,B,B^*$ is of the order $rd$ which is much smaller than than the size $(d-r)^2$ of $C$. Therefore, the asymptotic behaviour of the error matrix is determined by that of $C$ and its spectrum converges to the Wigner law in the limit of large sample size and large dimension.  In particular, the leading contribution to the norm error $\|\hat{\rho}_{\rm LS}-\rho_r\|$ for low rank states is $2\sqrt{d/N}$. A similar situation occurs for high rank states $r\approx d$ where the dominant variances are provided by the block $A$.

Indeed simulations for a pure and the fully mixed state of $n=7$ atoms with 
$N=10^6$ samples gave norm errors $0.0225$ and respectively $0.0221$, while the above theoretical prediction is $0.0226$. Further simulation results for $n=5$ and $n=6$ atoms states of different ranks show a good match with the above estimate.


{\it Trace norm risk.} 
As in the case of the operator norm, the trace norm error can be estimated thanks to the fact that the eigenvalues of error matrix follow an approximate Wigner semicircle distribution. Using
$$
\frac{1}{2\pi}\int_{-2}^2 |x| \sqrt{4-x^2} dx = \frac{8}{3\pi}
$$
we find that for low rank, or close to full rank states, the leading contribution to 
$\mathbb{E}(\|\hat{\rho}_{\rm LS} -\rho\|_1)$ is $8 d^{3/2}/ (3\pi\sqrt{N})$. 
This rate agrees with the upper bound in equation \eqref{eq.risk.ls.2design}, but provides an exact asymptotic constant for this rate. Indeed simulations with a pure and maximally mixed states of $n=7$ atoms and $N=10^6$ samples gave norm-one errors of $1.228$ and respectively 
$1.229$ while the theoretical prediction is $1.229$.



\section{Generalised least squares estimator} \label{sec.gls}

Consider the generic linear regression problem 
\begin{equation}\label{eq.regression}
{\bf y} = A {\bf x} + {\bf n}
\end{equation}
with ${\bf x}\in \mathbb{R}^a$ and unknown vector, ${\bf y}\in  \mathbb{R}^b$ a vector of observations and ${\bf n}$ the `noise' term with \emph{fixed and known} covariance matrix $C$, which is assumed to be strictly positive. While the LS estimator $\hat{\bf x}_{\bf LS}$ is optimal in the sense of minimising the \emph{prediction error} $\| A\hat{\bf x} - {\bf y}\|$, this is not the case when considering an \emph{estimation error} e.g. the mean square error $\mathbb{E} \|  \hat{\bf x} - {\bf x} \|^2$, unless the covariance matrix is proportional to the identity. 
To recover the `equal noise' situation we multiply equation \eqref{eq.regression} from the left by $C^{-1/2}$,  to obtain
$$
{\bf y}^\prime = A^\prime {\bf x}  + {\bf n}^\prime, \qquad {\rm Cov}({\bf n}^\prime)= {\rm Id}.
$$
The LS estimator of the last regression equation has the smallest covariance matrix among linear unbiased estimators 
\begin{equation*}
\hat{\bf x}_{\rm GLS} 
=\underset{{\bf x}^\prime}{\arg\min} \Vert {\bf y}^\prime - A^\prime {\bf x}^\prime \Vert^2 
= (A^TC^{-1}A)^{-1}A^T C^{-1} {\bf y}.
\label{eq1:GLSmin} 
\end{equation*}
This is called the generalised least squares (GLS) estimator.  when the noise distribution is Gaussian, the GLS estimator coincides with the MLE. In the i.i.d. setting where $m$ samples are available, the GLS estimator is asymptotically normal and efficient 
\begin{equation}
\sqrt{m}(\hat{\bf x}_{\rm GLS} - {\bf x}) \rightarrow 
N \left(\boldsymbol{0}, (A^T C^{-1} A)^{-1} \right).
\end{equation}

\emph{GLS as a projection.} Similarly to the LS estimator we consider the image
$\hat{{\bf y}}_{\rm GLS}= A \hat{\bf x}_{\rm GLS}$ of the GLS estimator  ${\bf x}_{\rm GLS}$. Then $\hat{{\bf y}}_{\rm GLS}$ is the projection of the data ${\bf y}$ onto the subspace ${\rm Ran}(A)\subset \mathbb{R}^b$ with respect to the covariance-dependent metric
$$
d_C({\bf y}, {\bf z}) = ({\bf y}- {\bf z})^T C^{-1} ({\bf y}- {\bf z}).
$$

\emph{GLS for tomography.} Let us return now to the tomography regression problem of the form given by equation \eqref{eq:matrixform}. While the LS estimator is optimal in the sense of equation \eqref{eq1:LSmin}, in general it is not optimal in the estimation sense for any locally quadratic distance function. To remedy this we would like to construct a corresponding generalised least square estimator to take into account the nontrivial form of the noise covariance matrix $ \Omega $ given by equation 
\eqref{eq1:covariance}. 
%
However, there are two issues which prevent us from directly applying the GLS methodology to the tomography data. The first is that $ \Omega$ is unknown since it depends on the true probabilities, and therefore on the unknown state $\rho$. 
We therefore propose to use an estimate of the covariance matrix instead. We describe the computation of this estimate in Appendix \ref{appendix.implement}.  The second difficulty is that $\Omega$ is a singular matrix due the constraint 
$\sum_{\boldsymbol{o}} f(\boldsymbol{o} \vert \boldsymbol{s}) =1$ for each setting ${\bf s}$. This means that for each setting ${\bf s}$ the vector 
$|{\bf 1}_{\bf s}\rangle\in \mathbb{C}^d$ is a zero eigenvector for $\Omega_{\bf s}$  and one should work within the orthogonal complement of $|{\bf 1}_{\bf s}\rangle$. This can be achieved  by choosing a block-diagonal isometry $V :\mathbb{C}^{k(d-1)} \to \mathbb{C}^{k\cdot d}$ such that each block $V_{\bf s} $ satisfies $ \langle{\bf 1}_s |V_{\bf s} =0$, and defining the new `frequencies' vector $\tilde{\boldsymbol{f}}$ with settings components 
$\tilde{\boldsymbol{f}}_{\bf s}  = V_{\bf s}^* \boldsymbol{f}_{\bf s}$. Similarly, we can also remove the trace-one constraint for states by choosing the vectorisation \eqref{eq1:paraLS} such that the last basis element is $\tau_{d^2} = \mathbb{1}/\sqrt{d}$. In this case the state is uniquely determined by the free parameters $\{\beta_i : 1\leq i \leq d^2-1\}$. If we denote by
$J :\mathbb{C}^{d^2-1}\to \mathbb{C}^{d^2}$ the isometric embedding with respect to the standard basis, then $\tilde{\boldsymbol{\beta}}:= J^* \boldsymbol{\beta}$ is the truncated vector of free parameters for $\rho$. 

The regression problem can now be written in terms of the `tilde' vectors and matrices 
$$
\tilde{\boldsymbol{f}}= \tilde{X} \tilde{\boldsymbol{\beta}} + \tilde{\boldsymbol{\epsilon}}
$$
where $\tilde{X} = V^* X J$ and $\tilde{\boldsymbol{\epsilon}}$ has covariance matrix
$\tilde{\Omega} = V^* \Omega V$ which is \emph{non-singular}.
We can now apply the GLS methodology and define the estimator as
\begin{equation*}
 \hat{\boldsymbol{\beta}}_{\rm GLS} = (\tilde{X}^* (\hat{\tilde{\Omega}})^{-1} \tilde{X})^{-1}\tilde{X}^* (\hat{\tilde{\Omega}})^{-1} {\boldsymbol{\tilde{f}}},
\end{equation*}
where $\hat{\tilde{\Omega}}$ is the (non-singular) estimated covariance matrix. We denote 
$\hat{\rho}_{\rm GLS}$ as estimate of the density matrix constructed from its vectorised form 
$\hat{\boldsymbol{\beta}}_{\rm GLS}$. For future use, let us denote $\tilde{\mathcal{M}}: M^1_{sa}(\mathbb{C}^d)\to \mathbb{R}^{k(d-1)}$ the map  $\rho\mapsto \tilde{X} \tilde{\boldsymbol{\beta}}$, so that $\hat{\tilde{p}}_{\rm GLS} =\tilde{\mathcal{M}} (\hat{\rho}_{\rm GLS}) $ .
%

\subsection{Asymptotic theory of GLS}\label{sec.gls.asymptotics}
What are the asymptotic properties of $\hat{\rho}_{\rm GLS}$? For large $m$, the noise distribution becomes Gaussian, and the covariance estimator converges to the actual covariance. The estimator $\hat{\boldsymbol{\beta}}_{\rm GLS}$ becomes asymptotically normal as $m\to\infty$
\begin{equation}
\sqrt{m}( \hat{\boldsymbol{\beta}}_{\rm GLS} - \boldsymbol{\beta}) \rightarrow 
N \left(\boldsymbol{0}, (\tilde{X}^* \tilde{\Omega}^{-1} \tilde{X})^{-1} \right).
\end{equation}
which means that the corresponding asymptotic average Fisher information per sample is 
$I_{\rm GLS}=\tilde{X}^* \tilde{\Omega}^{-1} \tilde{X}/k$. In appendix \ref{appendix.GLS} we show that $I_{\rm GLS}$ coincides with the Fisher information 
$I(\rho|\mathscr{S})$ defined as in equation \eqref{eq1:Fisher} for the parametrisation 
$\tilde{\boldsymbol{\beta}}$. This equality implies that the error rates of the GLS estimator have the same asymptotic behaviour as those of the uML estimator, and both estimators satisfy asymptotic normality. In fact one can make the stronger statement that the two estimators are asymptotically close to each other.  

\emph{Equivalence of GLS and uML.} As stated above, the GLS can be interpreted as the projection of the data $\tilde{\boldsymbol{f}}$ onto the image of $\tilde{X}$ in 
$\mathbb{R}^{k(d-1)}$ with respect to the metric
$$
d(\boldsymbol{g}, \boldsymbol{h}) = (\boldsymbol{g}-\boldsymbol{h})^T \hat{\tilde{\Omega}}^{-1}(\boldsymbol{g}-\boldsymbol{h})^T
$$ 
On the other hand, in section \ref{sec.mle.tomo} we showed that for large $N$, we can define the uML estimator as the projection of $\boldsymbol{f}$ onto the hyperplane $\mathcal{L}_d:= \mathcal{M} (M_{sa}^1(\mathbb{C}^d))$, with respect to the relative entropy distance. Since in the first order of approximation the relative entropy is given by the quadratic form $\tilde{\Omega}^{-1}$, the two projections become identical in the asymptotic limit.

\section{Thresholded least squares estimator } \label{sec1:TLS}

The LS estimator (as well as GLS) suffers from the disadvantage that it does not necessarily produce a density matrix, i.e, a positive semi-definite estimate of trace one. While the significant eigenvalues can be estimated reasonably well with enough data, the LS estimator will typically have negative eigenvalues corresponding to small or zero eigenvalues of the true state. The thresholded least squares estimator (TLS) proposed in \cite{spectralthresholding}, improves the LS estimator by selecting the state which is the closest in Frobenius norm to $\hat{\rho}_{\rm LS}$ and whose non-zero eigenvalues are above a certain threshold $\nu\geq 0$, i.e.
$$
\hat{\rho}_{\rm TLS} := \underset{\tau\in \mathcal{S}_d^{(\nu)}}{\arg\max} \|\hat{\rho}_{\rm LS} -\tau\|_2^2
$$
with $\mathcal{S}_d^{(\nu)} $ the set of density matrices with spectrum in $\{0\} \cup\{[\nu,1]\}$. The choice of the statistical noise threshold is informed by the accuracy of the LS estimate, and a theoretical and a `data-driven' choices for this threshold are detailed in \cite{spectralthresholding}, see also \cite{smolin_efficient_2012,Dryden2014}. In practice it is found that estimator's performance improves if the threshold is allowed to be `data-driven', by using cross-validation to choose the optimal value of the threshold, see Appendix \ref{appendix.implement}. 

It turns out that computing $\hat{\rho}_{\rm TLS}$ is very efficient and can be easily implemented.
Let 
$$
\hat{\rho}_{\rm LS}= \sum_{i=1}^d \hat{\lambda}_i |\hat{g}_i\rangle\langle \hat{g}_i|
$$
be the spectral decomposition of $\hat{\rho}_{\rm LS}$ where we assume that the eigenvalues are sorted in descending order $\hat{\lambda}_1 \geq \ldots \geq \hat{\lambda}_d$. The thresholded estimator 
$$
\hat{\rho}_{\rm TLS}= \sum_{i=1}^d \hat{\lambda}_i(\nu) |\hat{g}_i\rangle\langle \hat{g}_i|
$$
has the same eigenvectors as $\hat{\rho}_{\rm TLS}$, and its eigenvalues can be computed in terms of $\hat{\lambda}_i$ as summarised in Algorithm \ref{al1:TLS}.
 \begin{algorithm}
    \SetKwInOut{Input}{Input}
    \SetKwInOut{Output}{Output}

    \Input{ Eigenvalues of LS estimator $\hat{\lambda}_1 \geq \ldots \geq \hat{\lambda}_d$, and noise threshold $\nu$}
    \Output{Eigenvalues $\hat{\lambda}_1(\nu) \geq \ldots \geq \hat{\lambda}_d(\nu)$ of thresholded estimate $\hat{\rho}_{\rm TLS}$ }
    \For{$p =1, \ldots, d$}{
    \eIf{$\hat{\lambda}_{d-p+1} \geq \nu$}
      {
        STOP;
      }
      {
        $\hat{\lambda}_{d-p+1} \leftarrow 0$; \newline
        \For{$j=1,...,d-p$}{
        $\hat{\lambda}_j \leftarrow \hat{\lambda}_j + \frac{1}{d-p} \left( 1- \sum_{m=1}^{d-p} \hat{\lambda}_{m} \right)$
        }
      }
      }
    \caption{Algorithm to threshold the eigenvalues of the LS estimate}
    \label{al1:TLS}
\end{algorithm}
In words, the algorithm checks if the smallest eigenvalue of $\hat{\rho}_{\rm LS}$ is above the noise threshold, and if it is then 
$\hat{\rho}_{\rm TLS} = \hat{\rho}_{\rm LS}$ ( note that by construction ${\rm Tr} (\hat{\rho}_{\rm LS})=1$, so there is no need to renormalise the LS estimator).  On the other hand if the smallest eigenvalue is below the threshold, it is set to zero and the remaining eigenvalues are suitably shifted accordingly so that their sum is equal to one. The final estimate $\hat{\rho}_{\rm TLS}$ is constructed by replacing the eigenvalues of $\hat{\rho}_{\rm LS}$ with these thresholded eigenvalues $\hat{\lambda}_i(\nu)$. The theoretical properties of the TLS estimator are presented in \cite{spectralthresholding} and are similar to those of the projected least squares estimator which is discussed in more detail below.


\subsection{Thresholded Generalised Least Squares Estimator (TGLS)}

\noindent This estimator is obtained by using the GLS estimate $\hat{\rho}_{\rm GLS}$ instead of the LS estimate as a starting point for the thresholding procedure. The constant for thresholding is chosen in the same way by using cross-validation. The advantage of TGLS is that it starts from a more accurate estimator than LS, which leads to smaller errors compared to TLS. The price to pay is that it is somewhat more involved computationally, although still faster than ML.

\section{Positive least squares estimator } \label{sec.pos.ls}


 The positive least squares (posLS) estimator is obtained by restricting the minimisation in (\ref{eq1:LSmin}) to parameters  that correspond to density matrices $\tau \in  \mathcal{S}_d$. Let $\mathcal{M} : M(\mathbb{C}^d)  \mapsto \mathbb{C}^{kd}$ be the measurement map defined as $[\mathcal{M}(\tau)]_{\boldsymbol{o},\boldsymbol{s}} = {\rm Tr}\left[ \tau P^{\boldsymbol{s}}_{\boldsymbol{o}} \right]/|\mathcal{S}|$, 
Then, similarly to section \ref{sec.ls.inverse} we can express the posLS estimator as
%
%
\begin{equation}
\hat{\rho}_{\rm posLS} := \underset{\tau \in \mathcal{S}_d}{\arg \min} \Vert \mathcal{M}(\tau) - \boldsymbol{f} \Vert^2. 
\end{equation}
By comparing with \eqref{eq1:LSmin} we see that $\hat{\rho}_{\rm posLS}$ is the projection of 
$\hat{\rho}_{\rm LS}$ with respect to the distance on matrices induced by the euclidian distance on measurement probabilities $d_{\mathcal{M}}(\rho, \tau) = \|\mathcal{M}(\rho) - \mathcal{M}(\tau)\|$. To our knowledge, with the exception of the results in \cite{KalevBaldwin}, its theoretical properties have not been studied in detail. While its statistical performance greatly improves on that of the LS, 
the posLS estimator has the drawback that it cannot be expressed in a closed form, and its computational complexity is comparable to that of the ML estimator. 
 \subsection{Positive least squares estimator for covariant measurements}

Let us consider the special case of the covariant measurement defined in equation \eqref{eq.povm.covariant}. This measurement maps states into probability distributions in an (almost) isometric way \cite{GutaKahnKungTropp}
$$
\| \mathcal{M}(\rho) - \mathcal{M}(\tau)\|^2 = \frac{d}{d+1}\| \rho- \tau \|_2^2
$$
Indeed using Weingarten formulas \cite{Collins2006} we get for any traceless $X$
\begin{eqnarray*}
\| \mathcal{M}(X) \|^2 &=& d^2 \int {\rm d}P {\rm Tr} (PX)^2 =
d^2\int {\rm d} U  |\langle U1 | X |U 1\rangle|^2 \\
&=& d^2 \sum_{ijkl} X_{ij} \bar{X}_{kl} \int U_{j1} U_{k1} U^*_{1i} U^*_{1l}\, {\rm d} U 
= \frac{d}{d+1}\|X\|_2^2.
\end{eqnarray*}
As a corollary, we find that the posLS estimator is the projection of the LS estimator with respect to the Frobenius distance. Therefore, for this specific measurement, the posLS estimator coincides with the projected least squares estimator which will be discussed in section \ref{sec.pls}.

\subsection{Positive generalised least squares estimator }\label{sec.posGLS}

This estimator is defined in much the same way as the posLS estimator, by restricting the minimisation in (\ref{eq1:GLSmin}) to parameters that correspond to density matrices. In keeping with the discussion in section \ref{sec.gls}, we consider the truncated frequency vector 
$\tilde{\boldsymbol{f}}$. By analogy with the GLS estimator, the positive generalised least squares (posGLS) estimator is defined as
\begin{equation}\label{eq1:posGLS}
\hat{\rho}_{\rm posGLS} := \underset{\tau \in \mathcal{S}_d}{\arg \min} \left\| \hat{\tilde{\Omega}}^{-1/2} \left( \tilde{\mathcal{M}}(\tau) - \boldsymbol{\tilde{f}} \right) \right\|^2. 
\end{equation}
We will show that $\hat{\rho}_{\rm posGLS}$ is asymptotically equivalent to the ML estimator 
$\hat{\rho}_{\rm ML}$.

\emph{Asymptotic equivalence of posGLS and ML.} As in the case of GLS and uML, it is easier to work with the images in $\mathbb{C}^{k(d-1)}$ of the different estimators through the (injective) map $\tilde{\mathcal{M}}$; we denote by $\hat{\tilde{p}}_{\rm posGLS} =\tilde{\mathcal{M}}(\hat{\rho}_{\rm posGLS} )$ the probability vector corresponding to the posGLS estimator, and similarly for other estimators. Let us equip the the space of `frequencies' $\mathbb{C}^{k(d-1)}$ with the distance 
$$
d_{\Omega} (\tilde{ \boldsymbol{p}},\tilde{ \boldsymbol{q}}) =  
\left\|\tilde{\Omega}^{-1/2} \left( \tilde{\boldsymbol{p}} -\tilde{ \boldsymbol{q}}\right)\right\|^2 =
\left( \tilde{\boldsymbol{p}} -\tilde{ \boldsymbol{q}}\right)^T
\tilde{\Omega}^{-1}\left( \tilde{\boldsymbol{p}} -\tilde{ \boldsymbol{q}}\right)
$$
Therefore, $\hat{\tilde{p}}_{\rm posGLS}$ is the projection with respect to $d_{\Omega}$ of 
$\tilde{\bf f}$ onto the convex set 
$$
\mathcal{P}_d = \tilde{\mathcal{M}} (\mathcal{S}_d) \subset \tilde{\mathcal{M}} (M^1_{sa}(\mathbb{C}^d) ) \subset \mathbb{C}^{k(d-1)}.
$$  
On the other hand, the GLS estimator is the projection of $\tilde{\bf f}$ onto the hyperplane $\tilde{\mathcal{M}} (M^1_{sa}(\mathbb{C}^d))$ which contains $\mathcal{P}_d$. Therefore, by properties of projections on convex subsets we find that $\hat{\tilde{p}}_{\rm posGLS}$ is the projection of $\hat{\tilde{p}}_{\rm GLS}$ onto $\mathcal{P}_d$. 


In section \ref{sec.gls.asymptotics} we showed that the GLS estimator is asymptotically equivalent to the uML, as a consequence of the fact asymptotically with $m$ both projections are determined by the Fisher information metric. Since posGLS and ML are obtained by applying the same projections onto the smaller space $\mathcal{P}_d $, they are also asymptotically equivalent. Unfortunately, this equivalence does not provide us with  general estimation method which is more efficient than ML; the projection involved in posGLS does not seem to have closed form expression and requires a similar optimisation process as ML.

\section{Projected least squares}\label{sec.pls}
Recently, it has been shown that the theoretical properties $\hat{\rho}_{\rm TLS}$ are preserved even if the threshold $\nu$ is chosen to be zero \cite{GutaKahnKungTropp}. 
The \emph{projected least squares} (PLS) estimator is defined as
$$
\hat{\rho}_{\rm PLS} := \underset{\tau \in \mathcal{S}_d}{\arg \inf} \|\hat{\rho}_{\rm LS} -  \tau\|_2
$$
where the optimisation is performed over all states $\tau$. As in the case of the TLS estimator, this optimisation can be performed efficiently, and it only involves computing the spectral decomposition of $\hat{\rho}_{\rm LS}$ and applying Algorithm \ref{al1:TLS} with $\nu =0$.
The PLS estimator is therefore faster than the data-driven TLS and turns out to have quite similar behaviour to the latter. Note that in general PLS is different from posLS , as both can be seen as projections of the LS estimator with respect to different metrics. However, as noted before, the estimators coincide in the case of covariant measurements. In the next section we will study this scenario in more detail.

Using the LS concentration bound of Theorem \ref{th.LS}, the following \emph{rank dependent} norm-one bound for the PLS estimator was derived in \cite{GutaKahnKungTropp}, where the measurement is either a 2-design or the Pauli bases measurement.
\begin{theorem}\label{th.PLS}
Let $\hat\rho_{\rm PLS}$ be the PLS estimator of $\rho$ for a dataset consisting of $N=m \times k$ samples. 
Then 
\begin{equation*}
\mathrm{Pr} \left[ \left\| \hat{\rho}_{\rm PLS} - \rho \right\|_1 \geq \epsilon \right] \leq d 
\mathrm{e}^{ - \frac{N \epsilon^2}{43g(d) r^2} } \quad \tau \in \left[0,1 \right].
\end{equation*}
where $g(d)  =2 d $ for $2-$design measurements and $g(d) \simeq  d^{1.6}$ for Pauli measurements.
\end{theorem}
This gives a upper bound of $O( r^2 \cdot d \log d /\sqrt{N})$ on the convergence rate of norm-one for $2$-design measurements and $O(\log r^2 \cdot d^{1.6}\log d  /\sqrt{N})$ for Pauli measurements. In the first case the $\log d$ can be removed for covariant measurements and the resulting rate is optimal in the sense that it achives general lower bounds from \cite{HaahHarrow2017}.

\subsection{The asymptotic behaviour of PLS for covariant measurements}
\label{sec.PLS.covariant}


In this section we look in more detail at the PLS estimator in the context of covariant measurements defined in equation \eqref{eq.povm.covariant}. We have already seen that Theorem \ref{th.PLS} provides a concentration bound on the norm-one risk of the PLS estimator. While such results are very valuable thanks to their non-asymptotic nature, it is instructive and useful to also understand the asymptotic behaviour for large sample size $N$ and dimension $d$. Indeed, in section \ref{sec.asymptotics.ls.covariant} we showed how central limit  and random matrix theory can be used to obtain tighter bounds on estimation risks of the LS estimator.

We will be particularly interested in the behaviour of PLS for \emph{low rank states}. Due to covariance it suffices to choose states which are diagonal in the standard basis $\{|i\rangle: i=1, \dots, d\}$, and for simplicity we will restrict ourselves to rank-$r$ states with equal eigenvalues $\rho_r = \sum_{i=1}^r |i\rangle\langle i| /r$. As in section \ref{sec.asymptotics.ls.covariant} we write the PLS estimator as
\begin{equation}\label{eq.pls.block}
\hat{\rho}_{\rm PLS}  =
\rho_r+ 
 \frac{1}{\sqrt{N}}\left(
\begin{array}{cc}
\tilde{A}
& 
\tilde{B}
\\
\tilde{B}^*
& 
\tilde{C}
\end{array}
\right), \qquad 
\end{equation}
for some error blocks $\tilde{A},\tilde{B},\tilde{C}$ whose dependence on the $A,B,C$ blocks of the LS estimator needs to be determined.

Our analysis draws on the asymptotic theory of the LS described in section \ref{sec.asymptotics.ls.covariant}, and the arguments developed in \cite{BlumeKohout_2018} for the analysis of the ML estimator. 
We know that $\hat{\rho}_{\rm PLS}$ has the same eigenbasis as $\hat{\rho}_{\rm LS}$, and its eigenvalues are obtained from those of $\hat{\rho}_{\rm LS}$ by the truncation procedure which involves repeatedly setting to zero negative eigenvalues and shifting the remaining ones so that they add up to one. The following 3 step procedure aims to identify the leading contributions to PLS for low rank states, and $N\gg d\gg 1$.

\emph{1. Diagonalisation.} Recall that for large $N$ the LS estimator can be block diagonalised by means of a `small' unitary rotation $U$, cf. equation \eqref{eq.ls.rotation}
$$
U^*\hat{\rho}_{\rm LS}U
=
\left(
\begin{array}{cc}
I_r/r + A/\sqrt{N} & 0\\
0 & C/\sqrt{N}
\end{array}
\right) + o(N^{-1/2}).
$$ 
Therefore the eigenvalues of $\hat{\rho}_{\rm LS}$ can be grouped in two sets. The first is the set of eigenvalues of the block $C/\sqrt{N}$, which for large $d$ are distributed according to the Wigner semicircle law and lie between $\pm 2\sqrt{(d-r)/N}$.  The second set consists of the eigenvalues of the  block $I_r/r + A/\sqrt{N}$, which are small fluctuations of order $r/\sqrt{N}$ around $1/r$.

\emph{2. Truncation.}  As long as $N\gg d\gg 1$ and $r$ is small, the second set of eigenvalues is well separated from the first and it is very unlikely that any of these eigenvalues will be set to zero in the truncation process. Therefore, the cut-off point for the eigenvalues of $C/\sqrt{N}$ depends only on the sum of the larger eigenvalues
$$
 {\rm Tr} (I_r/r + A/\sqrt{N}) = 1 + {\rm Tr}(A)/\sqrt{N} =:1 +a/\sqrt{N}.
$$
 Moreover, since $(d-r)\gg 1$, the eigenvalues of $C/\sqrt{N}$ are (approximately) distributed according to the Wigner distribution 
\eqref{eq.wigner} on the interval $[-2\sqrt{(d-r)/N},2\sqrt{(d-r)/N} ]$. Therefore, we can write the 
cut-off point $q$ as the solution of the following equation \cite{BlumeKohout_2018}
$$
1= 1 +\frac{a}{\sqrt{N}}-rq + (d-r)\int_q^{2\sqrt{\frac{d-r}{N}}}  \, (x-q) \frac{N}{2\pi (d-r)} \sqrt{ \frac{4(d-r)}{N} - x^2} \, dx
$$
where the integral corresponds to the sum of the eigenvalues of the block 
$C/\sqrt{N}$ above $q$, after being shifted by $q$. Writing $q = 2\sqrt{d-r/N} \epsilon$ with $\epsilon = \epsilon(r,d,N,a)<1$, we get the following equation for $\epsilon$
\begin{equation}\label{eq.epsilon}
r \epsilon - \frac{a}{2\sqrt{d}} = \frac{2(d-r)}{\pi} \int_\epsilon^1 (y-\epsilon)\sqrt{1-y^2}dy.
\end{equation}
As the left side is smaller than $r$, the integral needs to be smaller than $r\pi/2(d-r)$, which means that $\epsilon$ is close to $1$ for $r\ll d$. This agrees with the intuition that a large part of the eigenvalues of the lower block will be set to zero by projecting the LS onto states. Further details on finding an (approximate) solution to \eqref{eq.epsilon} can be found in \cite{BlumeKohout_2018}. In particular, we will approximate $\epsilon$ by the \emph{deterministic} solution of equation \eqref{eq.epsilon} in which $a={\rm Tr}(A)$ is set to zero; indeed, for large $d$ this will have a negligible effect on $\epsilon$ but will allow us to compute 
$\epsilon$ it in terms of $r$ and $d$ deterministically. 

\emph{3. Rotation to original basis.}
Once the cut-off point $q$ has been computed, the projection of the rotated LS estimator $U^*\hat{\rho}_{\rm LS}U$ onto states can be written as 
$$
\left(
\begin{array}{cc}
\alpha I_r  + \frac{A}{\sqrt{N}} & 0\\
0 & \frac{{C}^\prime}{\sqrt{N}}
\end{array}
\right) + o\left(\frac{1}{\sqrt{N}}\right) ,
$$ 
where 
$$
\alpha:= \frac{1}{r} - 2\sqrt{\frac{d-r}{N}} \epsilon , \quad 
C^\prime = \left[C - 2 \sqrt{d-r} \epsilon I_{d-r}\right]_+ 
$$
and $[X]_+$ denotes the positive part of $X$. The PLS estimator is now obtained by performing the inverse rotation
\begin{eqnarray}
\hat{\rho}_{\rm PLS} &=& 
U 
\left(
\begin{array}{cc}
\alpha I_r + \frac{A}{\sqrt{N}} & 0\\
0 &\frac{C^\prime}{\sqrt{N}}
\end{array}
\right)
U^*  + o\left(\frac{1}{\sqrt{N}}\right)
\nonumber \\
&=& 
\rho_r + 
\frac{1}{\sqrt{N}} 
\left(
\begin{array}{cc}
A -2\sqrt{d-r}  \epsilon I_r &\alpha r B\\
\alpha r B^* &C^\prime
\end{array}
\right)+ o\left(\frac{1}{\sqrt{N}}\right)
\nonumber\\
&=& 
\rho_r + 
\frac{1}{\sqrt{N}} 
\left(
\begin{array}{cc}
\tilde{A} &\tilde{B}\\
\tilde{B}^* &\tilde{C}
\end{array}
\right)\label{eq.rho.pls.blocks}
\end{eqnarray}
We can now estimate the asymptotic risk of the PLS estimator with respect to the chosen loss functions.

{\it Frobenius risk.}
The mean square error scales as $N^{-1}$ and its rescaled version is
\begin{eqnarray*}
N\mathbb{E}\|\hat{\rho}_{\rm PLS} -\rho_r\|_2^2 &=& \mathbb{E}\|\tilde{A}\|_2^2 + 
2 \mathbb{E}\|\tilde{B}\|_2^2 + \mathbb{E}\|\tilde{C}\|_2^2
\end{eqnarray*}
The contribution from $\tilde{B}$ is 
$$
2\mathbb{E}\|\tilde{B}\|_2^2 = 2(\alpha r)^2 \mathbb{E} \|B \|_2^2
=
\frac{2(d+1)(d-r)(r+1)}{d+2} \left(1- 2r \sqrt{\frac{d-r}{N}} \epsilon\right)^2.
$$
where the variance of $B$ has been computed as in section \ref{sec.asymptotics.ls.covariant}; note that the term $2r\sqrt{(d-r)/N} \epsilon$ vanishes for $N\gg d$. This error is of the order $2rd$ and can be seen as stemming from the uncertainty in estimating the eigenbasis of $\rho_r$. For $\tilde{A}$ we write
$$
\mathbb{E} \|\tilde{A}\|_2^2 =\mathbb{E} \|A\|_2^2  + 4 r (d-r) \epsilon^2 = 
 (r+1)(r+2)\frac{d+1}{d+2} - \frac{(r+1)^2}{r} 
+ 4 r (d-r) \epsilon^2
$$
%
%
Finally, the term $ \mathbb{E}\|\tilde{C}\|_2^2$ is given by the sum of the squares of the remaining eigenvalues which can be approximated using the limit Wigner law as
\begin{eqnarray*}
\mathbb{E}\|\tilde{C}\|_2^2 &=& 
 N(d-r) \int_q^{2\sqrt{\frac{d-r}{N}}}  \, (x-q)^2 \frac{N}{2\pi (d-r)} \sqrt{ \frac{4(d-r)}{N} - x^2} \, dx \\
 &=& 
 \frac{8(d-r)^2}{\pi }\int_\epsilon^1 (y-\epsilon)^2 \sqrt{1-y^2} \, dy.
\end{eqnarray*}
Although the last term appears to be of order $(d-r)^2$, a careful analysis of the integral  \cite{BlumeKohout_2018} shows that it is of lower order than $rd$ due to the fact that $1-\epsilon$ leading term in a Taylor expansion is proportional to $(r/(d-r))^{2/5}$.
Therefore, by adding the three terms, we find that the Frobenius risk scales as $6 rd/N$. This agrees with the non-asymptotic results of \cite{GutaKahnKungTropp}, and provides the exact asymptotic constant of the Frobenius rate. By comparing with the lower bound to the asymptotic minimax rate of $2r(d-r)/N$ derived in \cite{spectralthresholding} we find that PLS is optimal for such states, within a constant which is at most 3. 

For the $n=7$ atoms state with $r=10$ the Frobenius error is $0.041$, compared to the theoretical one $0.043$. For rank $r=1$ state of 8 atoms with $N=10^5$ samples, the Frobenius error was $0.017$, while the theoretical prediction is $0.016$. The simulations results for all ranks of $n=5$ and $n=6$ atoms  are presented in panels a) and respectively b) of Figure \ref{fig.asymptotics.LS.PLS}.

{\it Operator norm risk.} 
Unlike the case of the LS estimator, the error matrix of PLS does not approach a Wigner distribution. For this reason, obtaining the asymptotic operator-norm risk seems difficult. However, the following lower bound follows from equation \eqref{eq.rho.pls.blocks}
$$
\sqrt{N}\|\hat{\rho}_{\rm PLS}-\rho_r\|\geq\max( \| \tilde{A}\| , \|\tilde{C} \| ) 
$$
Due to the truncation, the matrix $\tilde{C}$ is positive and the largest eigenvalue is approximtely 
$2\sqrt{d-r} (1-\epsilon)$. On the other hand, $\tilde{A}$ is dominated by the term $2\sqrt{d-r} \epsilon I_r$ whose norm is $2\sqrt{d-r} \epsilon$. Since $\epsilon\approx 1$ for large $d$, the lower bound is  $2\sqrt{d-r} \epsilon$. This complements the non-asymptotic upper bound of 
\cite{GutaKahnKungTropp} which has rate $O(\sqrt{d})$. Moreover, the lower bound seems to be a good approximation to the actual risk. A simulation with a rank $r=10$ state of $n=7$ atoms and 
$N=10^6$ samples gave an operator-norm error of $0.02$ while the lower bound is $0.01$.
For rank $r=1$ state of 8 atoms with $N=10^5$ samples, the operator-norm error was 
$0.12$, while the lower bound is $0.08$. 

%

{\it Trace-norm risk.} 
As in the case of the norm-error, we could not derive the asymptotic expression of the trace-norm risk but we can formulate a lower bound based on the pinching inequality
$$
\sqrt{N}\|\hat{\rho}_{\rm PLS}-\rho_r\|_1 \geq \|\tilde{A}\|_1 + \|\tilde{C} \|_1
$$
Note that $\tilde{A}=A- 2\sqrt{d-r}\epsilon I_r$, and the variance of the elements of $A$ is of the order 1. Therefore, for $r\ll d$ the shift $2\sqrt{d-r}$ dominates the eigenvalues of $A$ and $\tilde{A}$ is a negative matrix. In first approximation its norm-one is then $2r\sqrt{d-r}\epsilon$. On the other hand, $\tilde{C}$ is positive and its trace can be approximated as 
\begin{eqnarray*}
\mathbb{E}\|\tilde{C}\|_1 &=& 
 \sqrt{N}(d-r)\int_q^{2\sqrt{\frac{d-r}{N}}}  \, (x-q) \frac{N}{2\pi (d-r)} \sqrt{ \frac{4(d-r)}{N} - x^2} \, dx \\
 &=& 
 \frac{4(d-r)^{3/2}}{\pi }\int_\epsilon^1 (y-\epsilon) \sqrt{1-y^2} \, dy
\\
&=& 2r\sqrt{d-r}\epsilon
\end{eqnarray*}
where in last step we used equation \eqref{eq.epsilon} with $a=0$. Therefore the lower bound to the trace-norm error is $4r\sqrt{d-r}\epsilon$. For the $n=7$ atoms state with $r=10$ the trace-norm error was $0.33$ while the lower bound is  $0.21$. For rank $r=1$ state of 8 atoms with $N=10^5$ samples, the operator-norm error was $0.24$, while the lower bound is $0.16$.

{\it Bures risk.}
The Bures distance error can be expressed in terms of the blocks 
$\tilde{A}, \tilde{B}, \tilde{C}$ as follows
\begin{eqnarray}
d_B(\hat{\rho}_{\rm PLS}, \rho_r) 
&=& 2\left(1 - {\rm Tr} \left(\sqrt{\sqrt{\rho_r} \hat{\rho}_{\rm PLS} \sqrt{\rho_r}}\right)\right) 
= 
 2\left(1 - \frac{1}{\sqrt{r}}{\rm Tr} \left(\sqrt{I_r \hat{\rho}_{\rm PLS} I_r}\right)\right) \nonumber\\
 &=& 
 2\left(1 - \frac{1}{\sqrt{r}}{\rm Tr} \left(\sqrt{I_r/r + \tilde{A}/\sqrt{N}}\right)\right) =
 2\left(1 - \frac{1}{r}  {\rm Tr} \left(\sqrt{I_r + r \tilde{A}/\sqrt{N}}\right)\right) \nonumber\\
 &=& -\frac{1}{\sqrt{N}} {\rm Tr}(\tilde{A}) + \frac{r}{4N} {\rm Tr}(\tilde{A}^2) + o(N^{-1}) 
 \label{eq.bures.blocks}
\end{eqnarray}
where we used the Taylor expansion  in the last step. The leading term of the Bures risk is then
$$
-\frac{1}{\sqrt{N}}\mathbb{E} {\rm Tr}(\tilde{A})  =
\frac{2r \sqrt{d-r}}{\sqrt{N}} \epsilon (r,d) 
$$
where we used the fact that the LS block $A$ has centred distribution. The second order term
is 
\begin{eqnarray*}
\frac{r}{4N} \mathbb{E} {\rm Tr}(\tilde{A}^2) &=& 
\frac{r}{4N}  ( \mathbb{E} \|A\|^2 + 4r(d-r)\epsilon^2 )\\
&=&\frac{r}{4N} \left(  (r+1)(r+2)\frac{d+1}{d+2} - \frac{(r+1)^2}{r}  + 4r(d-r)\epsilon^2 \right)
\end{eqnarray*}
The simulation results for all ranks states of $n=5$ and $n=6$ atoms are presented panels c) and respectively d) of Figure \ref{fig.asymptotics.LS.PLS}.


\begin{figure}[h]
\begin{center}

\begin{minipage}[b]{0.45\linewidth}
\centering
\includegraphics[width=.9\linewidth]{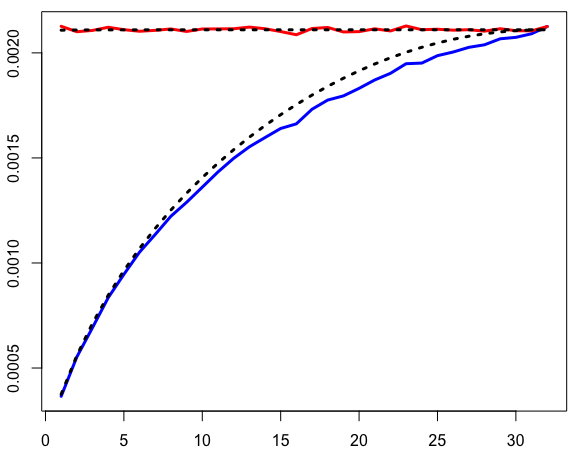}\\
a) Frobenius risks for $n=5$, $N=5\cdot 10^5$
 \end{minipage}
 \quad
 \begin{minipage}[b]{0.45\linewidth}
 \centering
\includegraphics[width=.9\linewidth]{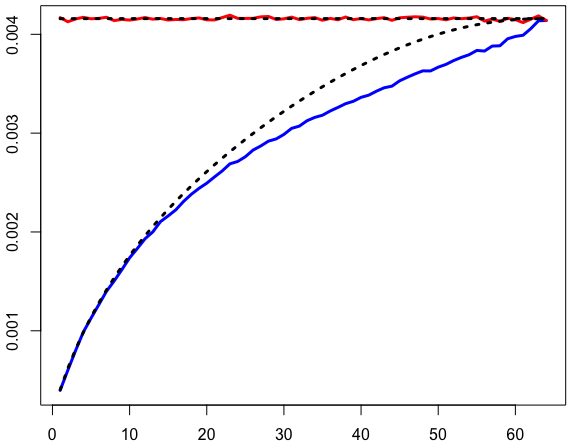}\\
b) Frobenius risks for  $n=6$, $N=10^6$ 
\end{minipage}
\vspace{4mm}

\begin{minipage}[b]{0.45\linewidth}
\centering
\includegraphics[width=0.9\linewidth]{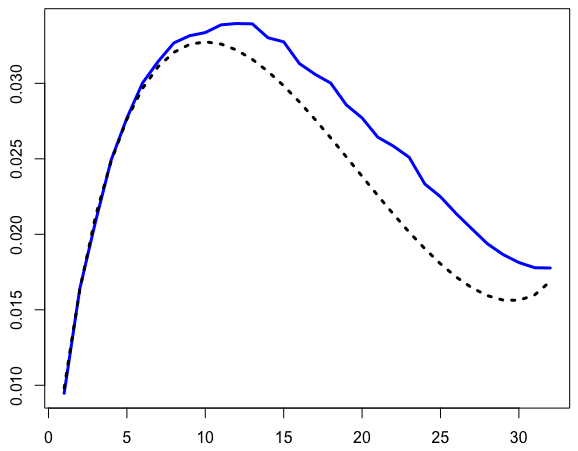}\\
c) Bures risks for $n=5$, $N=5\cdot 10^5$
 \end{minipage}
 \quad
 \begin{minipage}[b]{0.45\linewidth}
 \centering
\includegraphics[width=0.9\linewidth]{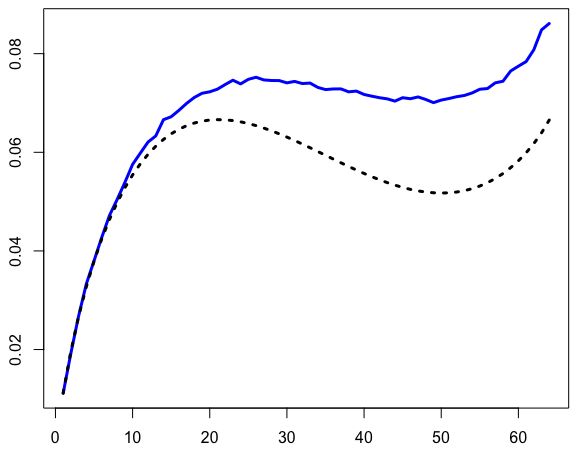}\\
d)  Bures risks for $n=6$, $N=10^6$ 
\end{minipage}

\end{center}
\caption{Frobenius and Bures risks for LS (red line) and PLS (blue line) versus the asymptotic predictions (black dotted lines)}
\label{fig.asymptotics.LS.PLS}

\end{figure}

\section{Comparative numerical simulations}\label{sec.simulations}

In this section we detail the methodology and results of a general simulation study which compares the performance of the estimators presented in the previous sections 
for a set of states 
and against several estimation criteria.
The states we consider are rank-$r$ states with a fixed spectrum of $r$ equal eigenvalues of magnitude $1/r$ each, and have randomly generated eigenvectors. 
This choice is motivated by the fact that such states are arguably harder to estimate among rank-r states (in analogy to the fact that an unbiased coin is harder to estimate than a biased one.
%
%
Additionally, having such a  spectrum allows for a more consistent comparison of the estimators across several ranks. 

We generate the above mentioned states for $3$ or $4$ qubits, and for a particular \enquote*{true state} we simulate a dataset $\mathcal{D}$ of counts from which the state is to be reconstructed. The outcome statistics depend on a few variables that we may vary, namely the type of measurement design (random basis vs Pauli), the number of repetitions per settings $m$, and in the case of the random basis measurements the total number $k$ of measured settings. This allows us to study the performance of the estimators across several different combinations of variables: types of states, ranks, measurement design, number of repetitions per setting $m$, the total number of settings $k$ and the number of ions $n$.
%
%
  
We will present plots of the estimated mean errors $\mathbb{E}\left[ D(\hat{\rho},\rho) \right]$ of the estimators for the equal eigenvalues states. 
For each given rank $r$ and number of qubits $n$, we generate a state with equal eigenvalues $\left( \frac{1}{r},\ldots,\frac{1}{r},0,\ldots,0 \right)$ and random eigenbasis. Then for each choice of measurement design and values of $k$ and $m$, the several estimates of the true state are evaluated. The error of each resulting estimate is computed using all the error functions listed in Table \ref{tb1:errorfunctions}, and the corresponding mean errors are estimated from 100 different runs of the experiment.

In order to make the results of the simulation study more accessible, we have made all plots for 3 and 4 atoms simulation available online via an interactive Rshiny application at this address:
\url{https://rudhacharya.shinyapps.io/plots/}, while a selection is presented in the paper.

\subsection{Squared Frobenius norm}

Figures \ref{fig1:frobeniuspauli} and \ref{fig1:frobeniusrandom4}  show the (estimated) Frobenius risk (mean square error) for states 4 qubits measured with the Pauli and the random basis design, respectively. The states are chosen randomly from the family of rank-$r$ states with equal non-zero eigenvalues. In both cases we note that the Frobenius risk of the LS estimator  has no significant dependence on the rank of the true state, and its performance is poor for small rank states. In contrast, the remaining estimators all show a scaling of the Frobenius risk with the rank of the true state. We also note that the performance of several of the estimators matches well with the Fisher-predicted risk. This is remarkable as the latter (\ref{eq1:MSE}) was defined for a rank-$r$ parameterisation of states, while none of the estimators have any prior knowledge of the rank.  

\begin{figure}[h]
\begin{center}
%

\begin{minipage}[b]{0.45\linewidth}
\centering
\includegraphics[width=\linewidth]{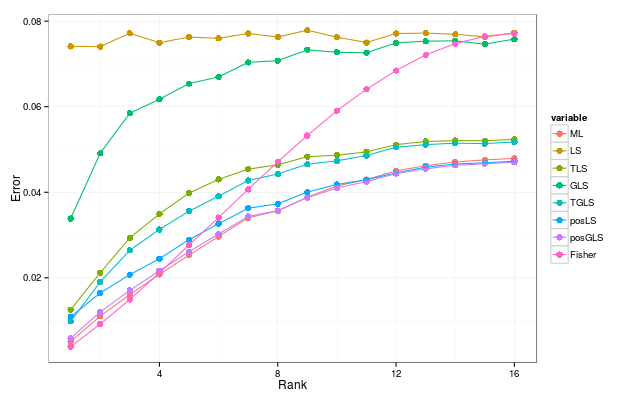}\\
a) $n=4$, $k=81$ and $m=100$ 
 \label{fig1:4_81_100}
 \end{minipage}
 \quad
 \begin{minipage}[b]{0.45\linewidth}
 \centering
\includegraphics[width=\linewidth]{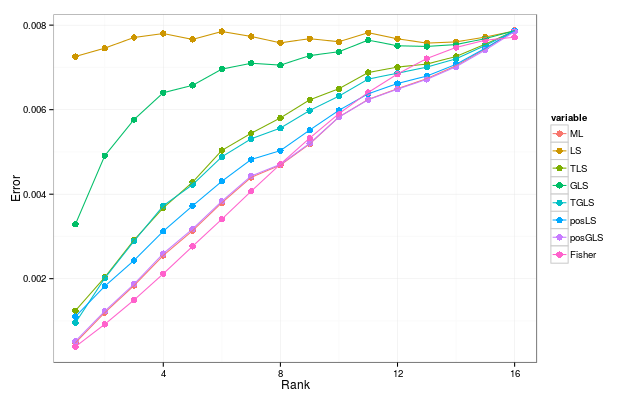}\\
b) $n=4$, $k=81$ and $m=1000$ 
\end{minipage}
\caption{The mean squared Frobenius error $\mathbb{E}\left[ \| \hat{\rho}-\rho\|_2^2 \right]$ of the estimators for random 4 qubit rank-$r$ states of equal eigenvalues, with Pauli measurements.}
\label{fig1:frobeniuspauli}
\end{center}
\end{figure}

\begin{figure}[t]
\begin{center}
%
%

\begin{minipage}[b]{0.45\linewidth}
\centering
\includegraphics[width=\linewidth]{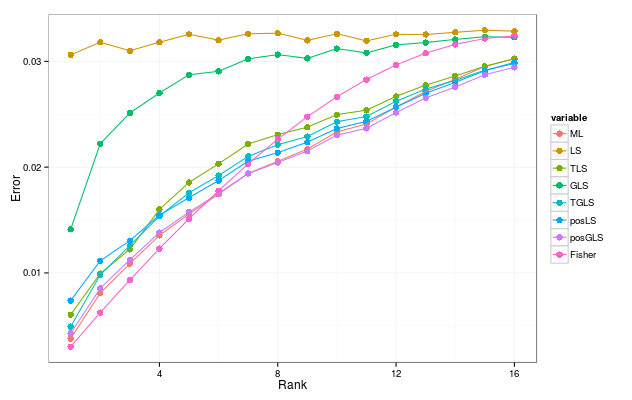}\\
a) $n=4$, $k=100$ and $m=100$ 
 \end{minipage}
 \quad
 \begin{minipage}[b]{0.45\linewidth}
 \centering
\includegraphics[width=\linewidth]{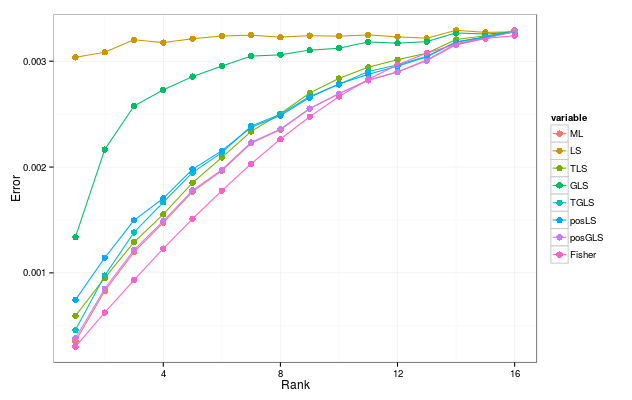}\\
b) $n=4$, $k=100$ and $m=1000$ 
\end{minipage}

\caption{The mean squared Frobenius error $\mathbb{E}\left[ \| \hat{\rho}-\rho\|_2^2 \right]$ of the estimators for 4 qubits  rank-$r$ states of equal eigenvalues with $k=100$ random bases and different repetition numbers. 
%
%
%
}
\label{fig1:frobeniusrandom4}

\end{center}

\end{figure}

We further note that for relatively small values of $N=m \times k$ the TLS, TGLS, posLS, posGLS, ML estimators significantly better that LS and GLS at higher ranks, while the errors approach each other for larger $N$, cf. Figures \ref{fig1:frobeniuspauli} a) and \ref{fig1:frobeniusrandom4} a) versus \ref{fig1:frobeniuspauli} b) and \ref{fig1:frobeniusrandom4} b). This reflects the fact that in the low $N$ case the asymptotic regime has not been reached and constrained estimators have an advantage even for full rank states which lie in the interior of the parameter space.
Indeed for an eigenvalue $\lambda$ of order $1/d$ and relatively small values of $N$, the unconstrained estimates $\hat{\lambda}$ will have a standard deviation of order 
$\sqrt{d/N}$ which may be comparable or larger than the magnitude of the eigenvalues themselves. Therefore without the constraint of positivity such estimators may produce estimates with $\hat{\lambda}<0$. In contrast, for the constrained estimators  the positivity constraint  provides additional information when the eigenvalues are small. 
 For large values of $N$ however, the uncertainty in the eigenvalues is very small and the Fisher risk acts as a lower bound for all of the estimators.

Across both the Pauli and the random measurement designs we note that the performance of the posGLS and the ML estimators is very similar, and for large $m$ almost identical. This confirms our asymptotic analysis which shows that the posGLS and the ML estimators are equivalent in the limit of large $m$, cf. section \ref{sec.posGLS}. 

\subsection{Bures and Hellinger distance}

As the Bures distance $D_{B}(\hat{\rho} ,\rho)^2$ is well defined only over density matrices, we plot the mean Bures errors only for the ML, TLS, TGLS, posLS , posGLS estimators. Figures \ref{fig1:burespauli} and \ref{fig1:buresrandom3} show the mean Bures errors for different sample sizes in the Pauli, and respectively random bases measurement design. For comparison, in Figures \ref{fig:mse.with.n} and \ref{fig:mf_3_random} we also plot the corresponding average Hellinger errors $D_H(\hat{\boldsymbol{\lambda}}, \boldsymbol{\lambda})^2$, cf. Table \ref{tb1:errorfunctions}. 

We note that the behaviour of the Hellinger errors is very similar to that of the Bures errors, with a better match for larger values of $N$. To give some intuition about this, we look at what happens in the case of qubits.  The Bures distance between neighbouring qubit states can be approximated by the sum of Hellinger distance and a quadratic form in the parameters of the unitary connecting the eigenbases of the two states \cite{AcharyaBures}
\begin{equation}\label{eq.Bures.approx}
 D_B(\rho,\hat{\rho})^2  \approx D_H(\boldsymbol{\lambda},\hat{\boldsymbol{\lambda}})^2  + \frac{1}{4} \frac{(1-2\lambda)(1-2\hat{\lambda})}{\sqrt{(1-\lambda)(1-\hat{\lambda})} 
+ \sqrt{\lambda \hat{\lambda}}}  \Phi^2 ,
\end{equation}
where $\Phi$ is the angle between the Bloch vectors of $\rho$ and $\hat{\rho}$. In a non-adaptive measurement scenario such as those considered here, the Bloch vector parameters can be estimated at rate $1/\sqrt{N}$ which means that the second term on the right side of  \eqref{eq.Bures.approx} is always of the order $1/N$. However, for states which are very pure 
($\lambda \approx 0$) the Hellinger component has the dominant contribution to the Bures distance, and is responsible for the ``non-standard'' scaling of $1/\sqrt{N}$ in the minimax risk \cite{AcharyaBures}. The simulation results indicate that a similar phenomenon may occur in higher dimensional systems. For full rank states, both the Bures and the Hellinger distance have a quadratic expansion, and in this sense a relation similar to \eqref{eq.Bures.approx} can be derived by splitting the Bures distance into quadratic contributions coming from changes in the eigenvalues and small basis rotations respectively, see also \cite{PereiraDelgado_Bures2018}. Alternatively, the block-matrix techniques used for analysing the Bures risk  in section \ref{sec.PLS.covariant} can be extended to rank deficient states of arbitrary spectrum to show that the leading $1/\sqrt{N}$ contribution comes from the Hellinger $D_H(\boldsymbol{\lambda},\hat{\boldsymbol{\lambda}})^2.$


Another noticeable feature across both the Pauli and the random bases measurement design is that for large $N$ the mean errors are seen to be larger for states of middling ranks than for the full rank states, see Figures 
\ref{fig1:burespauli}b, 
\ref{fig1:buresrandom3}}b. This is however not true for smaller values of $N$, as shown in Figures 
\ref{fig1:burespauli}a, 
\ref{fig1:buresrandom3}a. More precisely, for large $N$ we see a steep increase from pure states to low rank states, followed by an almost linear decrease down to the full rank state. A similar behaviour has been uncovered in the analysis of the PLS estimator for covariant measurements in section \ref{sec.PLS.covariant}. There, we found that for large $d$ and low rank $r$, the eigenvalues distribution of the error block $C$ of the LS estimator converges to the Wigner distribution, which allows us to compute the leading orders of the Bures risk of the PLS estimator. Since the covariant measurement arises in the large $m$ limit of the random basis measurement \cite{AcharyaGuta}, it is therefore expected that the risks  behave similarly in the two 
cases. Our simulations indicate that the mechanism governing the asymptotics of the Bures risk  seems to be robust with respect to the details of the measurement; although the Pauli basis measurement is not expected to produce an LS estimator whose error matrix is Wigner distributed, the Bures and Hellinger risks have similar characteristics as those of the covariant measurement. These findings are in line with those of \cite{BlumeKohout_2018}, and are worthy of further theoretical investigation.
\begin{figure}[h]
\begin{center}
%

\begin{minipage}[b]{0.45\linewidth}
\centering
\includegraphics[width=.9\linewidth]{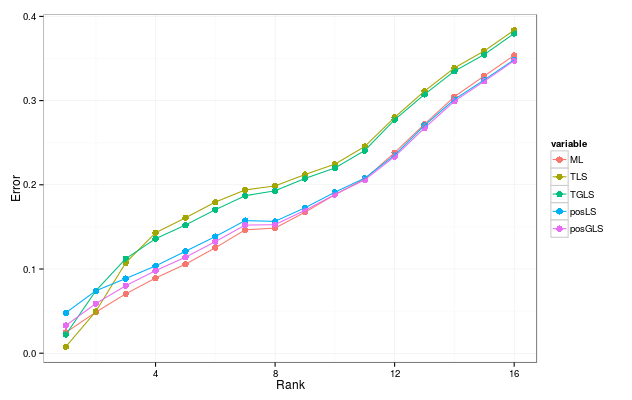}\\
a) $n=4$, $k=81$ and $m=100$ 
 \end{minipage}
 \quad
 \begin{minipage}[b]{0.45\linewidth}
 \centering
\includegraphics[width=.9\linewidth]{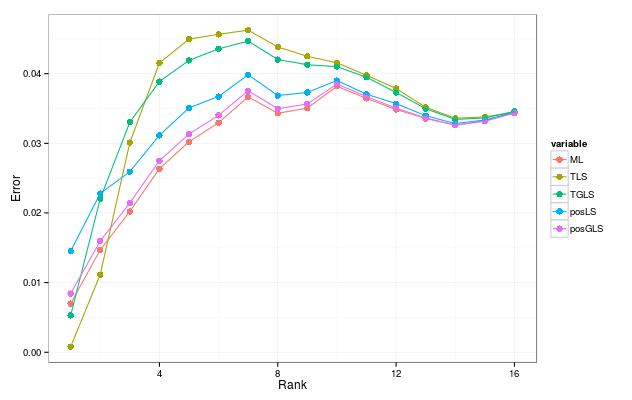}\\
b) $n=4$, $k=81$ and $m=1000$ 
\end{minipage}
\caption{The mean Bures error $\mathbb{E}\left[ D_{B}(\hat{\rho},\rho)^2 \right]$ of the estimators for random 4 qubits rank-$r$ states of equal eigenvalues, with Pauli measurement design.}
\label{fig1:burespauli}
\end{center}
\end{figure}

\begin{figure}[h!]
\begin{center}
%

\begin{minipage}[b]{0.45\linewidth}
\centering
\includegraphics[width=.9\linewidth]{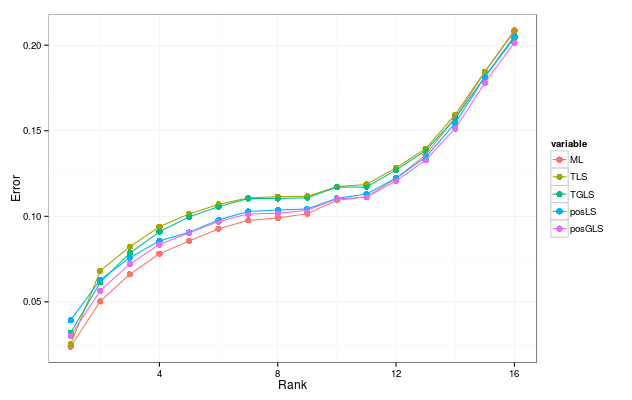}\\
a) $n=4$, $k=100$ and $m=100$ 
 \end{minipage}
 \quad
 \begin{minipage}[b]{0.45\linewidth}
 \centering
\includegraphics[width=.9\linewidth]{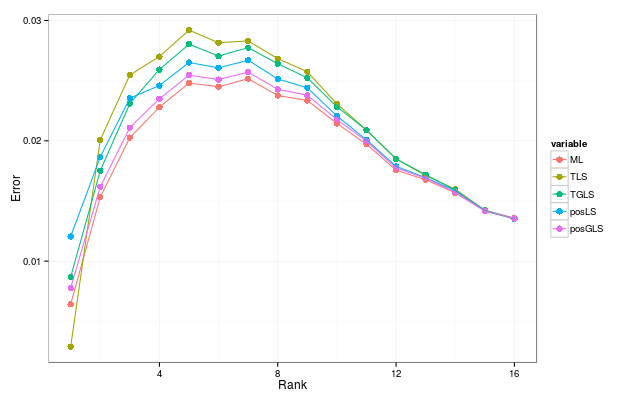}\\
b) $n=4$, $k=100$ and $m=1000$ 
\end{minipage}

%
\caption{The mean Bures error $\mathbb{E}\left[ D_{B}(\hat{\rho},\rho)^2 \right]$ of the estimators for random rank-$r$ states of equal eigenvalues, with random bases measurement design. 
}
\label{fig1:buresrandom3}
\end{center}
\end{figure}

While the Bures risk for rank deficient states scales as $1/\sqrt{N}$, for full rank states the Bures distance is locally quadratic, and standard asymptotic results imply that the convergence rate is in this case $1/N$, cf. section \ref{sec.mle.tomo}. In general,
 $$
D_B(\rho_{\boldsymbol{\theta} }, \rho_{\boldsymbol{\theta} +\delta\boldsymbol{\theta} } ) =
(\boldsymbol{\delta\theta})^{T} G_B(\boldsymbol{\theta}) (\boldsymbol{\delta\theta}) + O(\Vert \boldsymbol{\delta\theta} \Vert^2),
$$
with weight matrix $G_B(\boldsymbol{\theta}) =  F (\rho_{\boldsymbol{\theta}})/4$, where $F (\rho_{\boldsymbol{\theta}})$ is the quantum Fisher information. For the maximally mixed state 
$\rho_{r=d}$, and the parametrisation \eqref{eq1:parametrisation}, the latter has 2 uncorrelated blocks \cite{AcharyaGuta} 
corresponding to off-diagonal parameters 
$$
F_{a,b}  =2d \delta_{a,b}, \qquad 1 \leq a,b \leq d^2-d. 
$$
and respectively diagonal parameters
$$
F_{a,b}  =d(1+ \delta_{a,b}), \qquad d^2-d \leq a, b \leq  d^2-1 . 
$$
On the other hand, the classical average Fisher information for random basis measurements is given by $I = F/(d+1)$  \cite{spectralthresholding}. 
Since the ML estimator is asymptotically normal with variance $I(\boldsymbol{\theta})^{-1} /N$, we have
$$
N\mathbb{E} \left[ D_B(\hat{\rho}_{\rm ML}, \rho)^2\right] \to 
{\rm Tr} (G_B I^{-1} ) = \frac{(d^2-1)(d+1)}{4}.
$$  
For 3 qubits with $N=100\cdot 1000$ samples, and 4 qubits with $N=100\cdot 1000$, the asymptotic predictions for Bures errors are $0.0014$ and $0.01$ which match closely the simulations results \cite{simulations}. 
\begin{figure}[h]
\begin{center}
%

\begin{minipage}[b]{0.45\linewidth}
\centering
\includegraphics[width=.9\linewidth]{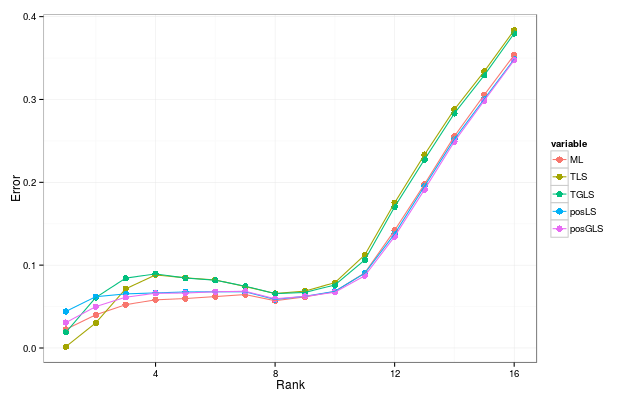}\\
a) $n=4$, $k=81$ and $m=100$ 
 \end{minipage}
 \quad
 \begin{minipage}[b]{0.45\linewidth}
 \centering
\includegraphics[width=.9\linewidth]{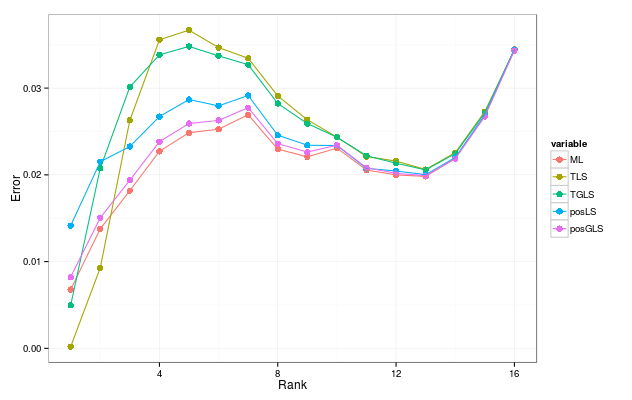}\\
b) $n=4$, $k=81$ and $m=1000$ 
\end{minipage}
\caption{The mean Hellinger error $\mathbb{E}\left[ D_{H}(\hat{\boldsymbol{\lambda}},\boldsymbol{\lambda})^2 \right]$ of the estimated eigenvalues $\hat{\boldsymbol{\lambda}}$ for random 4 qubits rank-$r$ states of equal eigenvalues, for Pauli measurement design.}
\label{fig:mse.with.n}
\end{center}
\end{figure}

\begin{figure}[h]
\begin{center}
%
\begin{minipage}[b]{0.45\linewidth}
\centering
\includegraphics[width=.9\linewidth]{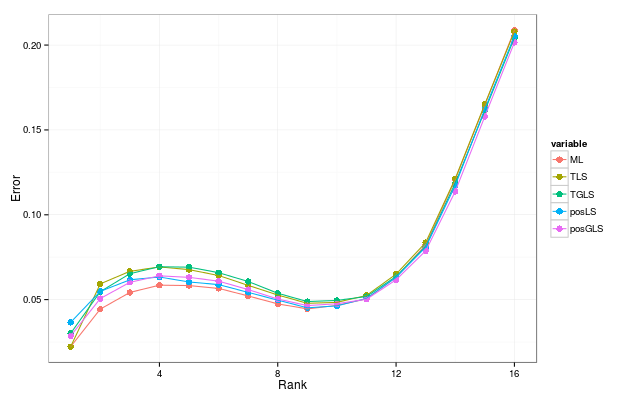}\\
a) $n=4$, $k=100$ and $m=100$ 
 \end{minipage}
 \quad
 \begin{minipage}[b]{0.45\linewidth}
 \centering
\includegraphics[width=.9\linewidth]{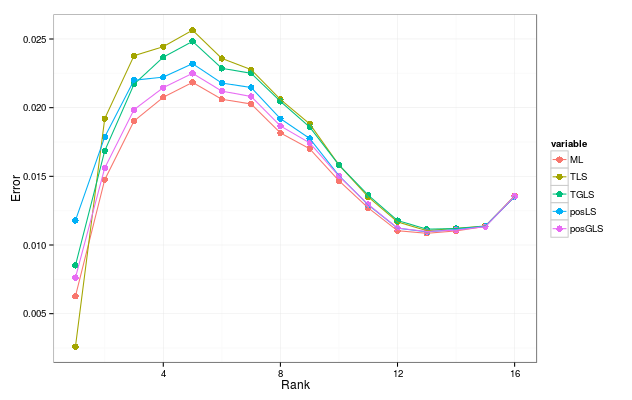}\\
b) $n=4$, $k=100$ and $m=1000$ 
\end{minipage}

\caption{The mean Hellinger error $\mathbb{E}\left[ D_{H}(\hat{\boldsymbol{\lambda}},\boldsymbol{\lambda})^2 \right]$ of the estimated eigenvalues $\hat{\boldsymbol{\lambda}}$ for random 4 qubits rank-$r$ states of equal eigenvalues, for random bases measurement design.}
 \label{fig:mf_3_random}
\end{center}
\end{figure}

%
%

\subsection{Trace-norm distance}

\begin{figure}[h]
\begin{center}
\begin{minipage}[b]{0.45\linewidth}
\centering
\includegraphics[width=.8\linewidth,height=0.55\linewidth]{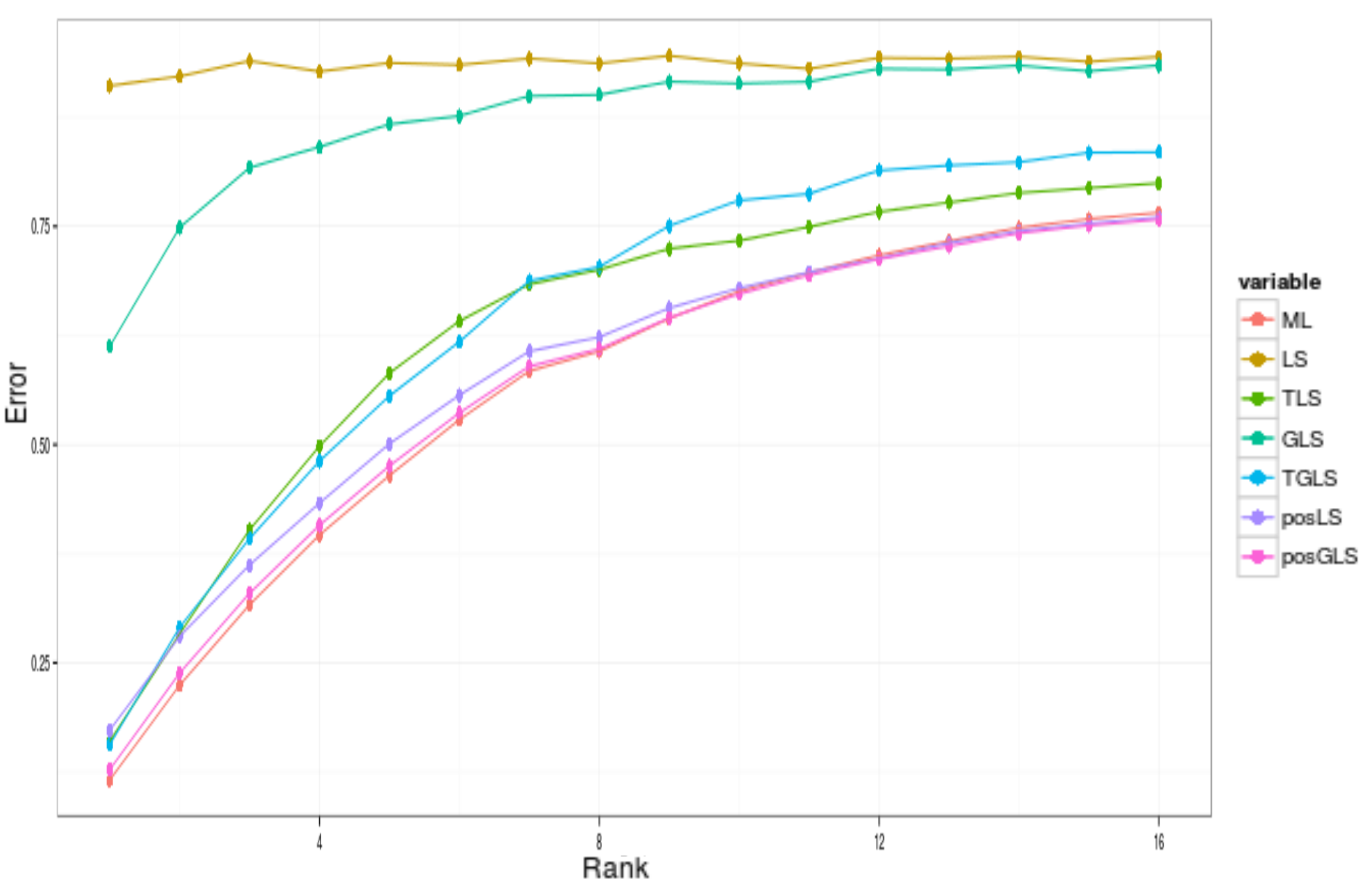}\\
a) $n=4$,  $m=100$ 
\end{minipage}
\quad
 \begin{minipage}[b]{0.45\linewidth}
 \centering
\includegraphics[width=.8\linewidth,height=0.55\linewidth]{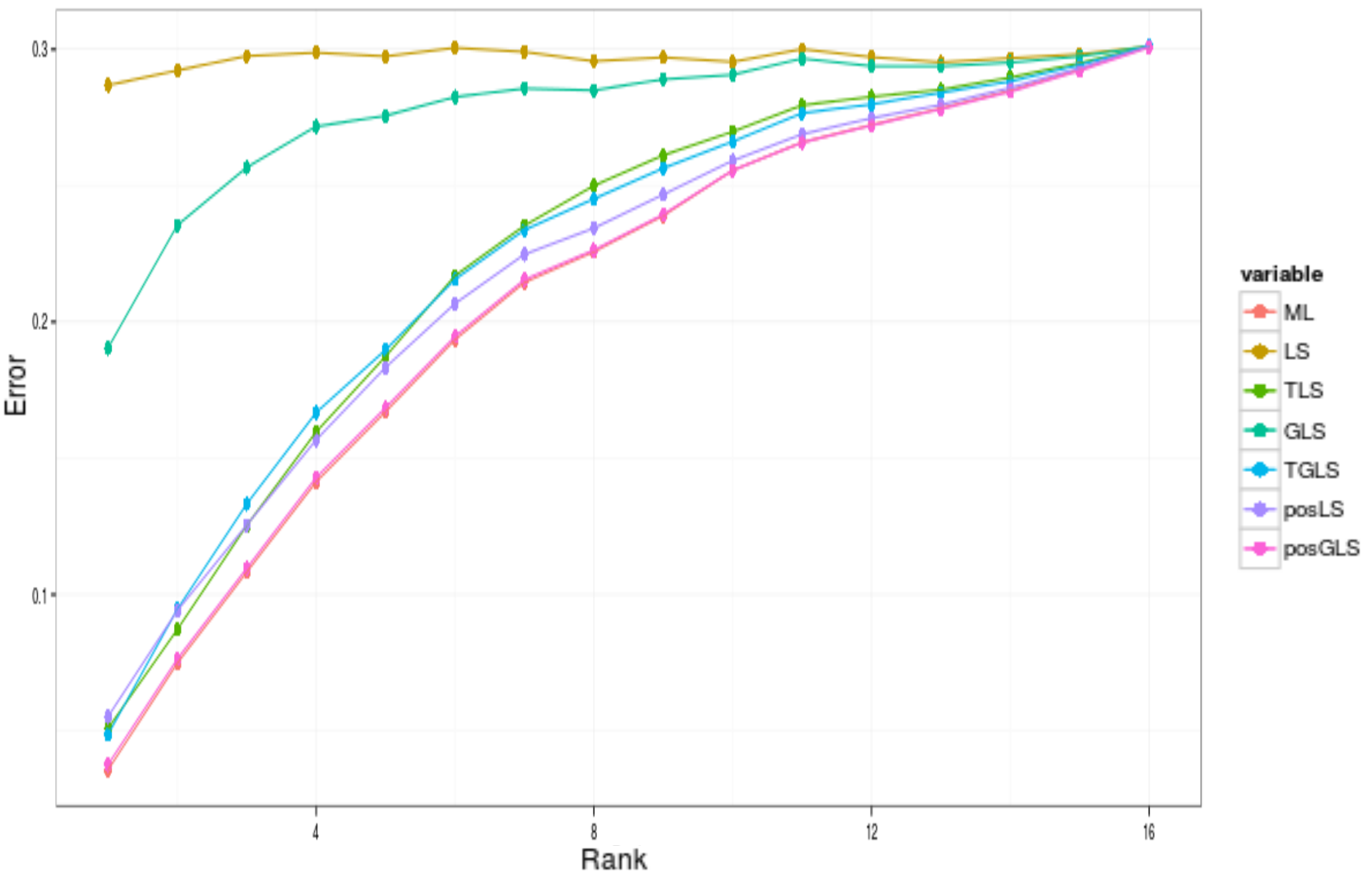}\\
b) $n=4$,  $m=1000$ 
\end{minipage}
%
\caption{The mean trace-norm error $\mathbb{E}\left[\| \hat{\rho}-\rho\|_1 \right]$ for Pauli bases measurements. Panels a) and b) show the risks for 4 qubits  rank-$r$ states of equal eigenvalues with different repetition numbers.}
\label{fig.L11}
\end{center}

\end{figure}

\begin{figure}[h]
\begin{center}
\begin{minipage}[b]{0.45\linewidth}
\centering
\includegraphics[width=.8\linewidth,height=0.55\linewidth]{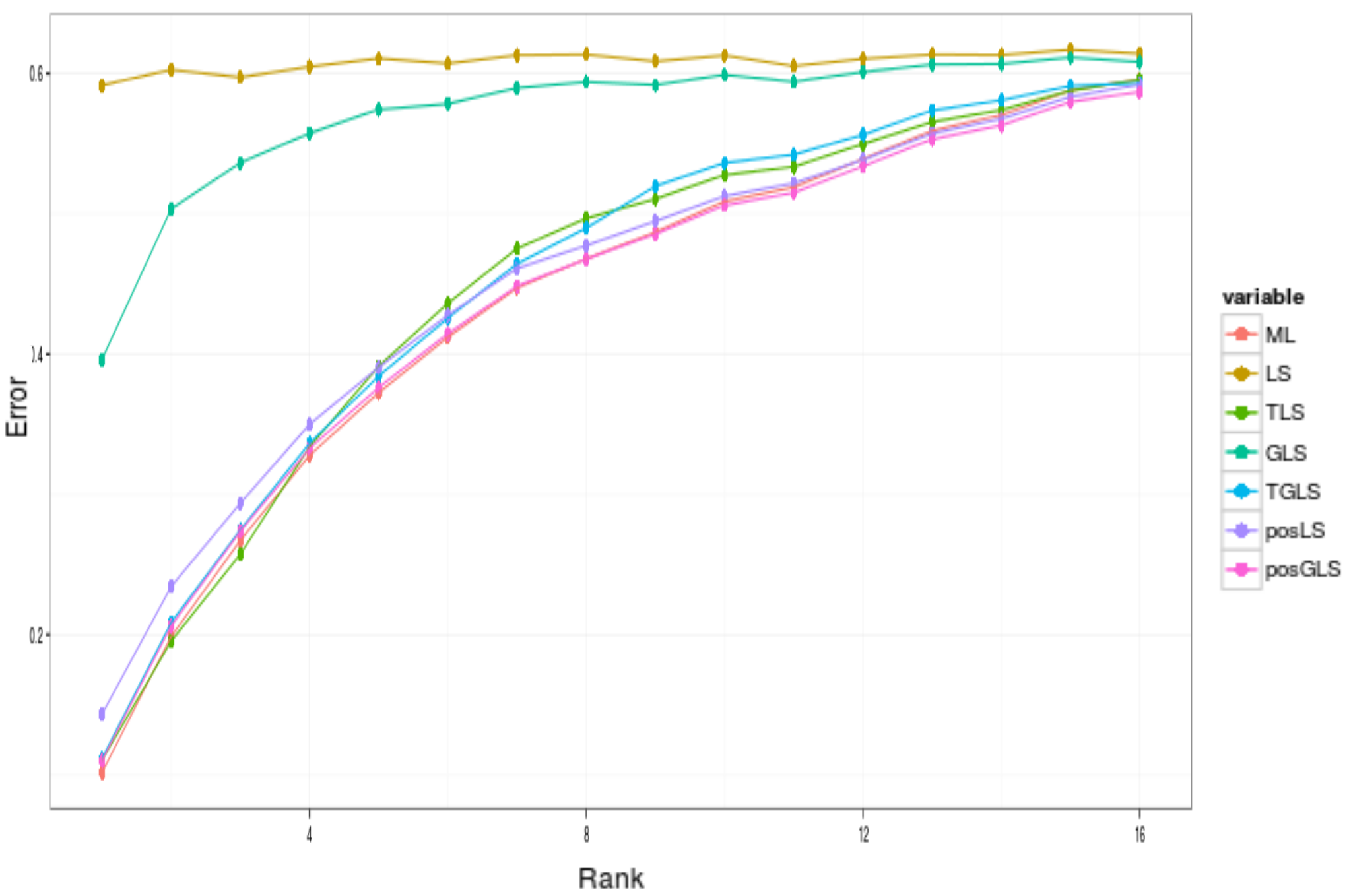}\\
a) $n=4$, $k=100$ and $m=100$ 
\end{minipage}
\quad
 \begin{minipage}[b]{0.45\linewidth}
 \centering
\includegraphics[width=.8\linewidth,height=0.55\linewidth]{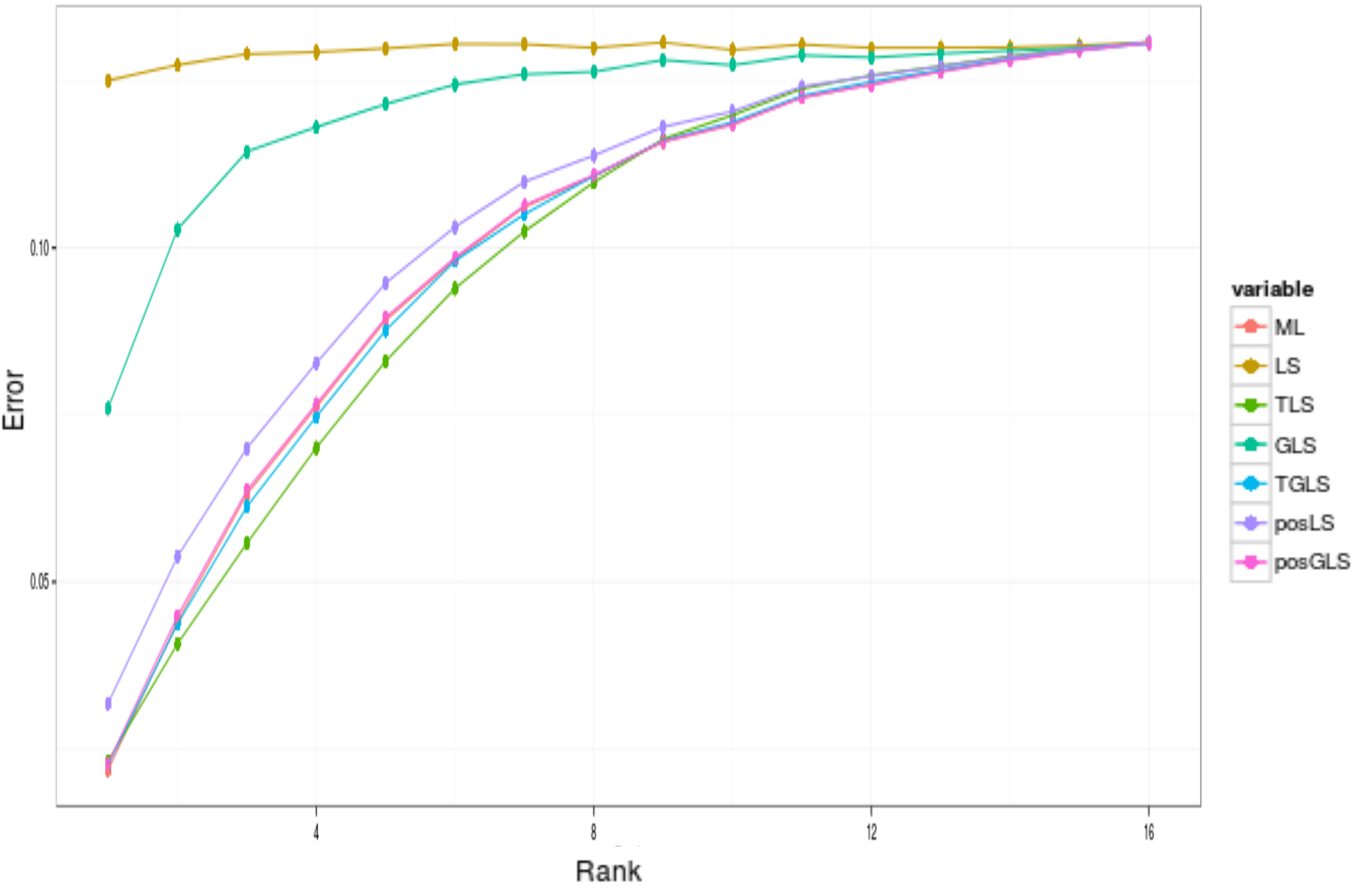}\\
b) $n=4$, $k=200$ and $m=1000$ 
\end{minipage}
%
\caption{The mean trace-norm error $\mathbb{E}\left[\| \hat{\rho}-\rho\|_1 \right]$ for random bases measurements. Panels a) and b) show the risks for 4 qubits  rank-$r$ states of equal eigenvalues with different number of bases and repetition numbers.}
\label{fig.L2}
\end{center}

\end{figure}

The risks for the trace-norm distance exhibit a similar behaviour to those of the Frobenius distance, as illustrated in Figures  \ref{fig.L11} and \ref{fig.L2}. A noticeable feature is that all constrained estimators have close risks for large sample sizes in both the Pauli and the random bases setting.

\section{Conclusions}

In this paper we studied theoretical an practical aspects of quantum tomography  methods. The unifying theme is that each estimator can be seen as a projection of the data onto a parameter space with respect to an appropriate metric. We considered estimators without positivity constraints (unconstrained maximum likelihood, least squares, generalised least squares) and with positivity constraints (maximum likelihood, positive least squares, thresholded least squares and projected least squares), and investigated the relationships between different estimators. While no estimator makes use of the state's rank (which is assumed to be unknown) the constrained estimators have significantly lower errors than the unconstrained estimators, for low rank states. To better understand this behaviour we derived new asymptotic error rates for the least squares estimator and for the projected least squares estimators, for a class of given rank states and covariant measurements. These results capture the exact rate dependence on rank and dimension and complement non-asymptotic concentration bounds of \cite{spectralthresholding,SugiyamaTurnerMurao,GutaKahnKungTropp}, showing that PLS has strong optimality properties; for instance the leading contribution to the Frobenius risk is $6rd$ which is `almost optimal', in that it is only 3 times larger than the minimax lower 
bound established in \cite{spectralthresholding}, which assumes that the rank is known. Our analysis has uncovered an interesting behaviour of the Bures error, which increases with respect to rank for small ranks and then decreases towards full rank states. The extensive simulations study shows that this behaviour (as well as the monotonic increase of other errors) is robust with respect to the measurement design. 

Computationally, maximum likelihood and positive least squares involve an optimisation over states and are significantly slower than projected least squares which only requires the diagonalisation of the least squares matrix followed by a simple truncation procedure. Our results confirm and strengthen those of \cite{GutaKahnKungTropp} and show that projected least squares is an attractive alternative to maximum likelihood, which is routinely used in practice. Additional improvements can be achieved by using generalised least squares as starting point, in a two steps procedure. 

An interesting and practically relevant open question is whether any of the `fast' estimators analysed here is statistically `optimal' for more realistic measurements such as the Pauli bases. More generally, one can ask whether these methods can be adapted to non-informationally complete measurement scenarios, and other physically motivated lower dimensional statistical models.


\emph{Acknowledgements.}{ We thank Jonas Kahn, Richard Kueng and Ian Dryden for fruitful discussions during the preparation of this manuscript.}

%
%
%
%
%
%

\appendix
\section{Appendix}\label{appendix}

\subsection{Fisher information for GLS}\label{appendix.GLS}

Here we prove the equality $I(\rho|\mathscr{S})  = I_{\rm GLS}$ stated in section \ref{sec.gls.asymptotics}.
Note that by definition
$$
I(\rho|\mathscr{S})  = \frac{1}{k}\tilde{X}^* {\rm Diag} (p_\rho^{-1}) \tilde{X} = 
 \frac{1}{k}J^* X ^* {\rm Diag} (p_\rho^{-1}) X J 
$$
and 
\begin{equation}\label{eq.covariance.equality1}
I_{\rm GLS} =\frac{1}{k} \tilde{X}^T \tilde{\Omega}^{-1} \tilde{X} = 
\frac{1}{k} J^* X^* V (V^* \Omega V)^{-1} V^* XJ.
\end{equation}
We will show that
\begin{equation}\label{eq.covariance.equality2}
V (V^* \Omega V)^{-1} V^* = {\rm Diag} (p_\rho^{-1}) + 
c|\mathbf{1}\rangle \langle \mathbf{1}| + d( |\mathbf{1}\rangle \langle p_\rho^{-1}| +  |p_\rho^{-1}\rangle \langle\mathbf{1}| )
\end{equation}
where $c, d$ are some constants and $|p_\rho^{-1}\rangle\in \mathbb{C}^{k\cdot d}$ is the vector whose entries are the inverses of measurement probabilities. Indeed, since $J^*X^* |\mathbf{1}\rangle =0$, equation \eqref{eq.covariance.equality2} implies  \eqref{eq.covariance.equality1}.

Now $V (V^* \Omega V)^{-1} V^*$ is the pseudo-inverse of $\Omega$ and since 
$V^*|\mathbf{1}\rangle =0$, it satisfies
$$
V (V^* \Omega  V)^{-1} V^* =  (\Omega - |\mathbf{1}\rangle \langle \mathbf{1}| )^{-1} + 
|\mathbf{1}\rangle \langle \mathbf{1}|
$$

To compute the inverse on the right side we use the definition \eqref{eq1:covariance} of $\Omega$ and apply the  Sherman-Morrison formula \cite{sherman1950} 
$$
(A-|x\rangle\langle x|)^{-1}= A^{-1}+ \frac{A^{-1}|x\rangle\langle x| A^{-1}}{1-\langle x|A^{-1}|x\rangle} 
$$ 
This gives
\begin{equation}\label{eq.b}
(\Omega - |\mathbf{1}\rangle \langle \mathbf{1}| )^{-1} =
\left({\rm Diag}(p_\rho) - |p_\rho\rangle\langle p_\rho | -  |\mathbf{1}\rangle \langle \mathbf{1}|\right)^{-1}=
B^{-1} + \frac{B^{-1}    |p_\rho\rangle\langle p_\rho | B^{-1}}{1- \langle p_\rho | B^{-1}|p_\rho \rangle}  
\end{equation}
where $B= {\rm Diag}(p_\rho) - |\mathbf{1}\rangle \langle \mathbf{1}|$. By applying the Sherman-Morrison formula again we get
$$
B^{-1} = {\rm Diag}(p_\rho^{-1}) +\frac{ |p_\rho^{-1}\rangle \langle p_\rho^{-1} |}{1- \langle \mathbf{1}| {\rm Diag}(p_\rho^{-1}) |\mathbf{1}\rangle } =
{\rm Diag}(p_\rho^{-1}) +\frac{ |p_\rho^{-1}\rangle \langle p_\rho^{-1} |}{1- q } 
$$
where $q= \sum_{{\bf o},{\bf s}} p^{-1}_\rho({\bf o}|{\bf s})$. Moreover, 
$$
B^{-1}    |p_\rho\rangle = |\mathbf{1}\rangle + \frac{kd }{1-q} |p^{-1}_\rho\rangle ,\qquad 
\langle p_\rho | B^{-1}|p_\rho \rangle = 1+ \frac{(kd)^2}{1-q}
$$
By plugging the last three expresssions into \eqref{eq.b} we get
\begin{eqnarray*}
(\Omega - |\mathbf{1}\rangle \langle \mathbf{1}| )^{-1}&=& 
{\rm Diag}(p^{-1}_\rho) +\frac{ |p_\rho^{-1}\rangle \langle p_\rho^{-1} |}{1- q } \\
&-& 
\frac{1-q}{(kd)^2} \left( |\mathbf{1}\rangle \langle \mathbf{1}| +  \frac{kd}{1-q} 
(|\mathbf{1}\rangle \langle p_\rho^{-1}| +  |p_\rho^{-1}\rangle \langle \mathbf{1}| )+ 
\frac{(kd)^2}{(1-q)^2} |p_\rho^{-1}\rangle \langle p_\rho^{-1}|\right)\\
&=&
{\rm Diag}(p^{-1}_\rho) + c|\mathbf{1}\rangle \langle \mathbf{1}| + d |\mathbf{1}\rangle \langle p_\rho^{-1}| + 
\bar{d} |p_\rho^{-1}\rangle \langle\mathbf{1}| 
\end{eqnarray*}
which concludes the proof of \eqref{eq.covariance.equality2} and of 
$I(\rho|\mathscr{S}) = I_{\rm GLS}$.

\subsection{Implementations of estimators}\label{appendix.implement}

Here we list certain practical details about the implementation of the estimators described in the paper. In particular we discuss how we compute the covariance matrix estimator $\hat{\tilde{\Omega}}$ used in the GLS, TGLS, posGLS estimators, and we also describe the cross-validation procedure we use to select a constant $C$ for the thresholded in the TLS and TGLS estimators.  

\begin{enumerate}

\item The covariance matrix estimator $\hat{\tilde{\Omega}}$ for the generalised estimators (GLS, TGLS and posGLS) is computed as  follows. Given a dataset $\mathcal{D}$, we first obtain the LS estimate $\hat{\rho}_{\rm LS}$ and then construct the TLS estimate (see Algorithm \ref{al1:TLS}) with threshold $\nu = 0$. From this we obtain an estimate of the probabilities $\hat{p}(\boldsymbol{o} \vert \boldsymbol{s}) = {\rm Tr}\left[ \hat{\rho}_{\rm TLS} P^{\boldsymbol{s}}_{\boldsymbol{o}} \right]$. The matrix $\hat{\tilde{\Omega}}$ is then constructed from these estimated probabilities via (\ref{eq1:covariance}). The generalised estimates (GLS/TGLS/posGLS) are then evaluated using $\hat{\tilde{\Omega}}$ and the same dataset $\mathcal{D}$.  \\


\item As mentioned briefly in section \ref{sec1:TLS},  the threshold for the TLS and TGLS estimators is selected using cross-validation. We describe this cross-validation method below \cite{spectralthresholding}. 
\begin{itemize}
\item For a particular number of repetitions per setting $m$, we simulate data in 5 independent batches $m/5$ repetitions per setting in each batch. and we denote the corresponding datasets as $\mathcal{D}_1, \ldots, \mathcal{D}_5$. The total dataset of counts is  the sum $\mathcal{D}= \sum_{i=1}^5 \mathcal{D}_i$. \\

\item We choose a vector of constants $C$ forming a mesh over the interval $[0,1]$. For each value of $C $, and for each $j \in \{1, \ldots, 5\}$ we compute the following estimators. We hold out the dataset $\mathcal{D}_j$, and compute the TLS/TGLS estimate $\hat{\rho}^{-j}_{\rm T(G)LS} (C)$ for the dataset $\mathcal{D}_{-j} = \sum_{i \neq j} \mathcal{D}_i$, with threshold $\nu =  C\sqrt{\frac{4}{m} \log{2^{n+1}}}$, cf. \cite{spectralthresholding}. For each $\mathcal{D}_j$ the LS estimate $\hat{\rho}_{\rm LS}^{j}$ is also evaluated.  \\

\item For all values of $C $, the empirical discrepancy is evaluated for a choice of error function $D(\hat{\rho}, \rho)$
\begin{equation*}
CV_{D}(C) = \frac{1}{5} \sum_{i=1}^5 D\left(\hat{\rho}^{-j}_{\rm T(G)LS} (C), \hat{\rho}^{j}_{\rm LS} \right)
\end{equation*}

\item This function $CV(C)$ is then minimised over all values $C$
\begin{equation*}
\hat{C}_{D} = \underset{C}{\arg \min}\, CV_{D}(C)
\end{equation*}

\noindent this gives an estimate for the holding constant, which is then used to evaluate the TLS  or the TGLS estimators with threshold $\nu = \hat{C}_D \sqrt{\frac{4}{m} \log{2^{n+1}}}$.

\end{itemize}
\noindent Notice that the cross-validation procedure picks different constants for different choices of the error function. An important caveat here is that the Bures distance is not defined for the LS estimates $\hat{\rho}^{j}_{\rm LS}$, and therefore the procedure above cannot apply. Instead in the simulations we estimate the thresholding constant $\hat{C}_{D_B}$ using the ML estimate as 
\begin{equation*}
\hat{C}_{D_{B}} = \underset{C}{\arg \min}\, D_{B}(\hat{\rho}_{\rm T(G)LS}, \hat{\rho}_{\rm ML})^2.
\end{equation*}

\end{enumerate}


\begin{thebibliography}{10}

\bibitem{simulations}
Complete simulation results.
\newblock \url{https://rudhacharya.shinyapps.io/plots/}.

\bibitem{cvx}
Cvx: Matlab software for disciplined convex programming.
\newblock http://cvxr.com/cvx/, 2017.

\bibitem{AcharyaGuta}
A.~Acharya and M.~Guta.
\newblock Statistical analysis of compressive low rank tomography with random
  measurements.
\newblock {\em Journal of Physics A: Mathematical and Theoretical},
  50(19):195301, 2017.

\bibitem{AcharyaBures}
A.~Acharya and M.~Guta.
\newblock Minimax estimation of qubit states with {Bures} risk.
\newblock {\em Journal of Physics A: Mathematical and Theoretical}, 51:175307,
  2018.

\bibitem{AcharyaTheoGuta}
A.~Acharya, T.~Kypraios, and M.~Guta.
\newblock Statistically efficient tomography of low rank states with incomplete
  measurements.
\newblock {\em New Journal of Physics}, 18(4):043018, 2016.

\bibitem{Gill&Guta&Artiles}
L.~Artiles, R.~D. Gill, and M.~Guta.
\newblock An invitation to quantum tomography.
\newblock {\em J. Royal Statist. Soc. B (Methodological)}, 67:109--134, 2005.

\bibitem{Audenaert_Scheel}
Koenraad M~R Audenaert and Stefan Scheel.
\newblock Quantum tomographic reconstruction with error bars: a kalman filter
  approach.
\newblock {\em New Journal of Physics}, 11(2):023028, 2009.

\bibitem{Bagan&Baig&Tapia}
E.~Bagan, M.~Baig, and R.~Munoz-Tapia.
\newblock Optimal scheme for estimating a pure qubit state via local
  measurements.
\newblock {\em Phys. Rev. Lett.}, 89:277904, 2002.

\bibitem{Bagan&Gill}
E.~Bagan, M.~A. Ballester, R.~D. Gill, A.~Monras, and R.~Munoz-Tapia.
\newblock Optimal full estimation of qubit mixed states.
\newblock {\em Phys. Rev. A}, 73:032301, 2006.

\bibitem{MLparis}
K.~Banaszek, G.~M. D'Ariano, M.~G.~A. Paris, and M.~F. Sacchi.
\newblock Maximum-likelihood estimation of the density matrix.
\newblock {\em Physical Review A}, 61, 1999.

\bibitem{Barndorff-Nielsen&Gill&Jupp}
O.~E. Barndorff-Nielsen, {Gill, R. D.}, and {Jupp, P.~E.}
\newblock On quantum statistical inference (with discussion).
\newblock {\em J. R. Statist. Soc. B}, 65:775--816, 2003.

\bibitem{BaumgratzCramerPlenio}
T.~Baumgratz, A.~N\"{u}{\ss}eler, M.~Cramer, and M.~B. Plenio.
\newblock A scalable maximum likelihood method for quantum state tomography.
\newblock {\em New Journal of Physics}, 15(12):125004, 2013.

\bibitem{Belavkin1976}
V.~P. Belavkin.
\newblock Generalized uncertainty relations and efficient measurements in
  quantum systems.
\newblock {\em Theoretical and Mathematical Physics}, 26:213--222, 1976.

\bibitem{MLrobin}
R.~Blume-Kohout.
\newblock Hedged maximum likelihood quantum state estimation.
\newblock {\em Phys. Rev. Lett.}, 105:200504, 2010.

\bibitem{BlumeKohout}
R.~Blume-Kohout.
\newblock Optimal, reliable estimation of quantum states.
\newblock {\em New Journal of Physics}, 12(4):043034, 2010.

\bibitem{BraunsteinCaves}
S.~L. Braunstein and C.~M. Caves.
\newblock Statistical distance and the geometry of quantum states.
\newblock {\em Phys. Rev. Lett.}, 72:3439--3443, May 1994.

\bibitem{spectralthresholding}
C.~Butucea, M.~Guta, and T.~Kypraios.
\newblock Spectral thresholding quantum tomography for low rank states.
\newblock {\em New Journal of Physics}, 17(11):113050, 2015.

\bibitem{christandl2012}
M.~Christandl and R.~Renner.
\newblock Reliable quantum state tomography.
\newblock {\em Phys. Rev. Lett.}, 109:120403, Sep 2012.

\bibitem{Collins2006}
B.~Collins and P.~{\'{S}}niady.
\newblock Integration with respect to the haar measure on unitary, orthogonal
  and symplectic group.
\newblock {\em Communications in Mathematical Physics}, 264:773--795, 2006.

\bibitem{Cramer:2010}
M.~Cramer, M.~B. Plenio, S.~T. Flammia, R.~Somma, D.~Gross, S.~D. Bartlett,
  O.~Landon-Cardinal, D.~Poulin, and Y.-K. Liu.
\newblock Efficient quantum state tomography.
\newblock {\em Nat Commun}, 1:149, 12 2010.

\bibitem{D'Ariano.0}
G.~M. D'Ariano, {Macchiavello, C.}, and {Paris, M.~G.~A.}
\newblock Detection of the density matrix through optical homodyne tomography
  without filtered back projection.
\newblock {\em Phys. Rev. A}, 50:4298--4302, 1994.

\bibitem{Dryden2014}
I.~L. Dryden, Le. H., S.~P. Preston, and A.~T.~A. Wood.
\newblock Mean shapes, projections and intrinsic limiting distributions.
\newblock {\em Journal of Statistical Planning and Inference}, 145:25--32,
  2014.

\bibitem{Erdos}
Laszlo Erd\"{o}s.
\newblock Universality of wigner random matrices: a survey of recent results.
\newblock {\em Russian Mathematical Surveys}, 66(3):507, 2011.

\bibitem{FaistRenner}
P.~Faist and R.~Renner.
\newblock Practical and reliable error bars in quantum tomography.
\newblock {\em Phys. Rev. Lett.}, 117:010404, Jul 2016.

\bibitem{FerrieBK}
C.~Ferrie and R.~Blume-Kohout.
\newblock Minimax quantum tomography: Estimators and relative entropy bounds.
\newblock {\em Phys. Rev. Lett.}, 116:090407, 2016.

\bibitem{Ferrie2014}
Christopher Ferrie.
\newblock Self-guided quantum tomography.
\newblock {\em Phys. Rev. Lett.}, 113:190404, Nov 2014.

\bibitem{CSerrorbounds}
S.~T. Flammia, D.~Gross, Y.-K. Liu, and J.~Eisert.
\newblock Quantum tomography via compressed sensing: error bounds, sample
  complexity and efficient estimators.
\newblock {\em New Journal of Physics}, 14:095022, 2012.

\bibitem{Fujiwara&Nagaoka}
A.~Fujiwara and H.~Nagaoka.
\newblock Quantum fisher metric and estimation for pure state models.
\newblock {\em Phys. Lett A}, 201:119--124, 1995.

\bibitem{tenqubit2010}
W.-B. Gao, C.-Y. Lu, X.-C. Yao, P.~Xu, O.~Guhne, A.~Goebel, Y.-A. Chen, C.-Z.
  Peng, Z.-B. Chen, and J.-W. Pan.
\newblock Experimental demonstration of a hyper-entangled ten-qubit schrodinger
  cat state.
\newblock {\em Nat Phys}, pages 331--335, 2010.

\bibitem{GillGuta2013}
R.~D. Gill and M.~Guta.
\newblock On asymptotic quantum statistical inference.
\newblock {\em From Probability to Statistics and Back: High-Dimensional Models
  and Processes}, 9(10.1214/12-IMSCOLL909):105--127, 2013.

\bibitem{Gill&Massar}
R.~D. Gill and S.~Massar.
\newblock State estimation for large ensembles.
\newblock {\em Phys. Rev. A}, 61:042312, 2000.

\bibitem{LocalML}
D.~S. Goncalves, M.~A. Gomes-Ruggiero, C.~Lavor, O.~J. Farias, and P.~H.~S.
  Ribeiro.
\newblock Local solutions of maximum likelihood estimation in quantum state
  tomography.
\newblock {\em arXiv:1103.3682v2}, 2012.

\bibitem{GranadeCombes}
C.~Granade, J.~Combes, and D.~G. Corrie.
\newblock Practical bayesian tomography.
\newblock {\em new Journal of Physics}, 18:033024, 2016.

\bibitem{GranadeFerrieFlammia2017}
C.~Granade, C.~Ferrie, and S.~T. Flammia.
\newblock Practical adaptive quantum tomography.
\newblock {\em New Journal of Physics}, 19(11):113017, 2017.

\bibitem{CSnoRIP}
D.~Gross, Y.-K. Liu, S.~T. Flammia, S.~Becker, and J.~Eisert.
\newblock Quantum state tomography via compressed sensing.
\newblock {\em Phys. Rev. Lett.}, 105:150401, 2010.

\bibitem{GutaKahnKungTropp}
M.~Guta, J.~Kahn, R.~Kueng, and J.~A. Tropp.
\newblock Fast state tomography with optimal error bounds.
\newblock arXiv:1809.11162, 2018.

\bibitem{HaahHarrow2017}
J~Haah, A.~W. Harrow, Z.~Ji, X.~Wu, and N.~Yu.
\newblock Sample-optimal tomography of quantum states.
\newblock {\em IEEE Transactions on Information Theory}, 63:5628--5641, 2017.

\bibitem{Haffner2005}
H.~Haffner, W.~Hansel, C.~F. Roos, J.~Benhelm, D.~Chek-al kar, M.~Chwalla,
  T.~Korber, U.~D. Rapol, M.~Riebe, P.~O. Schmidt, C.~Becher, O.~Guhne, W.~Dur,
  and R.~Blatt.
\newblock Scalable multiparticle entanglement of trapped ions.
\newblock {\em Nature}, pages 643--646, 2005.

\bibitem{HannemannReiss}
Th. Hannemann, D.~Reiss, Ch. Balzer, W.~Neuhauser, P.~E. Toschek, and Ch.
  Wunderlich.
\newblock Self-learning estimation of quantum states.
\newblock {\em Phys. Rev. A}, 65:050303, May 2002.

\bibitem{Hayashi.editor}
M.~Hayashi, editor.
\newblock {\em Asymptotic theory of quantum statistical inference: selected
  papers}.
\newblock World Scientific, 2005.

\bibitem{Hayashi&Matsumoto}
M.~Hayashi and K.~Matsumoto.
\newblock Asymptotic performance of optimal state estimation in qubit system.
\newblock {\em J. Math. Phys.}, 49:102101--33, 2008.

\bibitem{Helstrom69}
C.~W. Helstrom.
\newblock {Quantum detection and estimation theory}.
\newblock {\em Journal of Statistical Physics}, 1:231--252, 1969.

\bibitem{Holevo}
A.~S. Holevo.
\newblock {\em Probabilistic and Statistical Aspects of Quantum Theory},
  volume~1.
\newblock Edizioni della Normale, 1 edition, 2011.

\bibitem{MLmofqubits}
D.~F.~V. James, P.~G. Kwiat, W.~J. Munro, and A.~G. White.
\newblock Measurement of qubits.
\newblock {\em Phys. Rev. A}, 64:052312, 2001.

\bibitem{KahnGuta2009}
J.~Kahn and M.~Guta.
\newblock Local asymptotic normality for finite dimensional quantum systems.
\newblock {\em Communications in Mathematical Physics}, 289(2):597--652, Jul
  2009.

\bibitem{KalevBaldwin}
A.~Kalev and C.~H. Baldwin.
\newblock The power of being positive: Robust state estimation made possible by
  quantum mechanics.
\newblock {\em arXiv preprint arXiv:1511.01433}, November 2015.

\bibitem{Keyl&Werner}
M.~Keyl and R.~F. Werner.
\newblock Estimating the spectrum of a density operator.
\newblock {\em Phys. Rev. A}, 64:052311, 2001.

\bibitem{KoltchinskiiXia}
V.~Koltchinskii and D.~Xia.
\newblock Optimal estimation of low rank density matrices.
\newblock {\em Journal of Machine Learning Research}, 16:1757--1792, 2015.

\bibitem{kueng_low_2017}
R.~Kueng, H.~Rauhut, and U.~Terstiege.
\newblock Low rank matrix recovery from rank one measurements.
\newblock {\em Appl. Comput. Harmon. Anal.}, 42(1):88 -- 116, 2017.

\bibitem{LanyonMaier}
B.~P. Lanyon, C.~Maier, M.~Holz\"{a}pfel, T.~Baumgratz, C.~Hempel, P.~Jurcevic,
  I.~Dhand, A.~S. Buyskikh, A.~J. Daley, M.~Cramer, M.~B. Plenio, R.~Blatt, and
  C.~F. Roos.
\newblock Efficient tomography of a quantum many-body system.
\newblock {\em Nature Physics}, 13:1158--1162, 2017.

\bibitem{LehmannCasella}
E.~L. Lehmann and G.~Casella.
\newblock {\em Theory of point estimation}.
\newblock Springer, 1998.

\bibitem{Leonhardt}
U.~Leonhardt.
\newblock {\em Essential Quantum Optics From Quantum Measurements to Black
  Holes}.
\newblock Number 9780521869782. Cambridge University Press, 1 edition, March
  2010.

\bibitem{LiEnglert2016}
X.~Li, J.~Shang, H.~K. Ng, and B.-G. Englert.
\newblock Optimal error intervals for properties of the quantum state.
\newblock {\em Phys. Rev. A}, 94:062112, Dec 2016.

\bibitem{Liu2011}
Y.{-}K. Liu.
\newblock Universal low-rank matrix recovery from pauli measurements.
\newblock In {\em Advances in Neural Information Processing Systems 24: 25th
  Annual Conference on Neural Information Processing Systems 2011. Proceedings
  of a meeting held 12-14 December 2011, Granada, Spain.}, pages 1638--1646,
  2011.

\bibitem{MahlerRozema}
D.~H. Mahler, L.~A. Rozema, A.~Darabi, C.~Ferrie, R.~Blume-Kohout, and A.~M.
  Steinberg.
\newblock Adaptive quantum state tomography improves accuracy quadratically.
\newblock {\em Phys. Rev. Lett.}, 111:183601, Oct 2013.

\bibitem{Massar&Popescu}
S.~Massar and S~Popescu.
\newblock Optimal extraction of information from finite quantum ensembles.
\newblock {\em Phys. Rev. Lett.}, 74:1259--1263, 1995.

\bibitem{Matsumoto}
K.~Matsumoto.
\newblock A new approach to the cramer-rao type bound of the pure state model.
\newblock {\em J. Phys. A}, 35(13):3111--3123, 2002.

\bibitem{Monz2011}
T.~Monz, P.~Schindler, J.~T. Barreiro, M.~Chwalla, D.~Nigg, W.~A. Coish,
  M.~Harlander, W.~H\"ansel, M.~Hennrich, and R.~Blatt.
\newblock 14-qubit entanglement: Creation and coherence.
\newblock {\em Phys. Rev. Lett.}, 106:130506, 2011.

\bibitem{ParisRehacek}
M.~G.~A. Paris and J.~Rehacek, editors.
\newblock {\em Quantum state estimation}.
\newblock Number 649 in Lecture Notes in Physics. Springer, Berlin Heidelberg,
  2004.

\bibitem{PereiraDelgado2018}
L.~Pereira, L.~Zambrano, J.~Cort\'es-Vega, S.~Niklitschek, and A.~Delgado.
\newblock Adaptive quantum tomography in high dimensions.
\newblock {\em Phys. Rev. A}, 98:012339, 2018.

\bibitem{PereiraDelgado_Bures2018}
L.~Pereira, L.~Zambrano, J.~Cort\'es-Vega, S.~Niklitschek, and A.~Delgado.
\newblock Adaptive quantum tomography in high dimensions.
\newblock {\em Phys. Rev. A}, 98:012339, 2018.

\bibitem{Petz&Jencova}
D.~Petz and {Jencova, A.}
\newblock Sufficiency in quantum statistical inference.
\newblock {\em Commun. Math. Phys.}, 263:259 -- 276, 2006.

\bibitem{QiBo}
B.~Qi, Z.~Hou, L.~Li, D.~Dong, G.~Xiang, and G.~Guo.
\newblock Quantum state tomography via linear regression estimation.
\newblock {\em Scientific Reports}, 3:3496 EP --, 2013.

\bibitem{HradilML}
J.~Rehacek, Z.~Hradil, E.~Knill, and A.~I. Lvovsky.
\newblock Diluted maximum-likelihood algorithm for quantum tomography.
\newblock {\em Phys. Rev. A}, 75:042108, 2007.

\bibitem{SchenkerSchulz-Baldes_dependent_2005}
J.~H. Schenker and H.~Schulz-Baldes.
\newblock Semicircle law and freeness for random matrices with symmetries or
  correlations.
\newblock {\em Mathematical Research Letters}, 12:531--542, 2005.

\bibitem{BlumeKohout_2018}
T.~L. Scholten and R.~Blume-Kohout.
\newblock Behavior of the maximum likelihood in quantum state tomography.
\newblock {\em New Journal of Physics}, 023050:20, 2018.

\bibitem{SchwemmerToth}
C.~Schwemmer, G.~T\'oth, A.~Niggebaum, T.~Moroder, D~Gross, O.~G\"uhne, and
  H.~Weinfurter.
\newblock Experimental comparison of efficient tomography schemes for a
  six-qubit state.
\newblock {\em Phys. Rev. Lett.}, 113:040503, Jul 2014.

\bibitem{sherman1950}
J.~Sherman and W.~J. Morrison.
\newblock Adjustment of an inverse matrix corresponding to a change in one
  element of a given matrix.
\newblock {\em Ann. Math. Statist.}, 21:124--127, 1950.

\bibitem{smolin_efficient_2012}
John~A. Smolin, Jay~M. Gambetta, and Graeme Smith.
\newblock Efficient method for computing the maximum-likelihood quantum state
  from measurements with additive gaussian noise.
\newblock {\em Phys. Rev. Lett.}, 108:070502, Feb 2012.

\bibitem{SteffensRio}
A.~Steffens, C.~A. Riofr{\'\i}o, W.~McCutcheon, I~Roth, B.~A. Bell,
  A.~McMillan, M.~S. Tame, J.~G. Rarity, and J.~Eisert.
\newblock Experimentally exploring compressed sensing quantum tomography.
\newblock {\em Quantum Science and Technology}, 2(2):025005, 2017.

\bibitem{SuessGross}
D.~Suess, L.~Rudnicki, T.~O. Maciel, and D.~Gross.
\newblock Error regions in quantum state tomography: computational complexity
  caused by geometry of quantum states.
\newblock {\em New Journal of Physics}, 19(9):093013, 2017.

\bibitem{SugiyamaTurnerMurao}
T.~Sugiyama, P.~S. Turner, and M.~Murao.
\newblock Precision-guaranteed quantum tomography.
\newblock {\em Phys. Rev. Lett.}, 111:160406, 2013.

\bibitem{TobiasToth}
Moroder. T., P.~Hyllus, G.~T{\'o}th, C.~Schwemmer, A.~Niggebaum, S.~Gaile,
  O.~G{\"u}hne, and H.~Weinfurter.
\newblock Permutationally invariant state reconstruction.
\newblock {\em New Journal of Physics}, 14(10):105001, 2012.

\bibitem{Hradilincomplete}
Yong~Siah Teo, Berthold-Georg Englert, Jaroslav {\v R}eh{\'a}{\v c}ek, Zden{\v
  e}k Hradil, and Dmitry Mogilevtsev.
\newblock Verification of state and entanglement with incomplete tomography.
\newblock {\em New Journal of Physics}, 14:105020, 2012.

\bibitem{Englert2011}
Yong~Siah Teo, Huangjun Zhu, Berthold-Georg Englert, Jaroslav \ifmmode
  \check{R}\else \v{R}\fi{}eh\'a\ifmmode~\check{c}\else \v{c}\fi{}ek, and
  Zden\ifmmode \check{e}\else~\v{e}\fi{}k Hradil.
\newblock Quantum-state reconstruction by maximizing likelihood and entropy.
\newblock {\em Phys. Rev. Lett.}, 107:020404, 2011.

\bibitem{tropp_user-friendly_2012}
J.~A. Tropp.
\newblock User-friendly tail bounds for sums of random matrices.
\newblock {\em Found. Comput. Math.}, 12(4):389--434, 2012.

\bibitem{Vidal}
G.~Vidal, J.~I. Latorre, P.~Pascual, and R.~Tarrach.
\newblock Optimal minimal measurements of mixed states.
\newblock {\em Phys. Rev. A}, 60:126, 1999.

\bibitem{YoungSmith}
G.~A. Young and R.~L. Smith.
\newblock {\em Essentials of statistical inference}.
\newblock Cambridge University Press, 2005.

\bibitem{Yuen&Lax}
H.~P. Yuen and M.~Lax.
\newblock Multiple-parameter quantum estimation and measurement of
  non-selfadjoint observables.
\newblock {\em IEEE Trans. Inform. Theory}, 19:740, 1973.

\end{thebibliography}

\end{document}